\documentclass{article}
\usepackage{amsmath}
\usepackage{graphicx}
\usepackage{amssymb}
\usepackage{subfigure}
\usepackage{cancel}
\usepackage{cite}
\usepackage{enumerate}
\usepackage[top=1in, bottom=1in, left=0.8in, right=0.8in]{geometry}

\newcommand{\beq}{\begin{equation}}
\newcommand{\eeq}{\end{equation}}
\newcommand{\bea}{\begin{eqnarray}}
\newcommand{\eea}{\end{eqnarray}}
\newcommand{\m}{\mathfrak{h}}

\newcommand{\gsim}{\lower.7ex\hbox{$\;\stackrel{\textstyle>}{\sim}\;$}}
\newcommand{\lsim}{\lower.7ex\hbox{$\;\stackrel{\textstyle<}{\sim}\;$}}

\newcommand{\bi}{\begin{itemize}}
\newcommand{\ei}{\end{itemize}}

\usepackage{color}
\usepackage{hyperref} 
\hypersetup{linktocpage=true}
\usepackage[all]{hypcap}
\hypersetup{
    bookmarks=true,         
    unicode=false,          
    pdftoolbar=true,        
    pdfmenubar=true,        
    pdffitwindow=false,     
    pdfstartview={FitH},    
    pdftitle={My title},    
    pdfauthor={Author},     
    pdfsubject={Subject},   
    pdfcreator={Creator},   
    pdfproducer={Producer}, 
    pdfkeywords={keyword1} {key2} {key3}, 
    pdfnewwindow=true,      
    colorlinks=true,       
    linkcolor=blue,          
    citecolor=magenta,        
    filecolor=magenta,      
    urlcolor=cyan           
}

\usepackage{fancyhdr}
\begin{document}
\thispagestyle{empty}
\vspace*{-22mm}
\begin{flushright}
UND-HEP-11-BIG\hspace*{.08em}04\\
\end{flushright}
\vspace*{10mm}

\vspace*{10mm}

\begin{center}
{\Large {\bf\boldmath
On CP Asymmetries  in Two-, Three- and Four-Body $D$ Decays}}
\vspace*{10mm}

{\bf  Ikaros\,I.\ Bigi and Ayan\ Paul} \\
\vspace{4mm}
{\small
{\sl Department of Physics, University of Notre Dame du Lac}\\
{\sl Notre Dame, IN 46556, USA}\vspace{1mm}
}


\vspace*{10mm}

{\bf Abstract}\vspace*{-1.5mm}\\
\end{center}

Indirect and direct CP violations have been established in $K_L$ and $B_d$ decays. They have been found 
in two-body decay channels -- with the exception of $K_L\to \pi^+\pi^-e^+e^-$ transitions. Evidence for direct 
CP asymmetry has just appeared in LHCb data on $A_{\rm CP}(D^0 \to K^+K^-) - A_{\rm CP}(D^0 \to \pi^+\pi^-)$ 
with 3.5 $\sigma$ significance. Manifestations of New Dynamics (ND) can appear in CP asymmetries 
just below experimental bounds. We discuss $D^{\pm}_{(s)}$, 
$D^0/\bar D^0$ and $D_L/D_S$ transitions to 2-, 3- and 4-body final states with a comment on 
predictions for inclusive vs. exclusive CP asymmetries. In particular we discuss 
T asymmetries in $D\to h_1h_2l^+l^-$ in analogy with $K_L\to \pi^+\pi^-e^+e^-$ transitions 
due to interference between M1, internal bremsstrahlung and possible E1 amplitudes. Such an effect 
depends on the strength of CP violation originating from the ND -- 
as discussed here for Little Higgs Models with T parity and non-minimal Higgs sectors -- 
but also in the interferences between these amplitudes even in the Standard Model. 
More general lessons can be learnt for T asymmetries in non-leptonic $D$ decays like 
$D \to h_1h_2 h_3 h_4$. Such manifestations of ND can be tested at LHCb and other 
Super-Flavour Factories like the projects at KEK near Tokyo and at Tor Vergata/Frascati near Rome.

\noindent
\vspace{2cm}

\begin{center}
{\em Dedicated to a great physicist and a wonderful friend -- Bruce Winstein (1943-2011)}
\end{center}

\vspace*{10mm}
\newpage
\thispagestyle{fancy}
\fancyhf{}

\lhead{\fancyplain{}{\nouppercase{}}}
\cfoot{\fancyplain{}{}}
\tableofcontents
\begin{flushleft}
{\bf\line(1,0){500}}
\end{flushleft}

\newpage

\setcounter{page}{1}

\fancyhf{}
 
\lhead{1\;\;\;\; Introduction}
\cfoot{\thepage}

\section{Introduction}

The Cabibbo-Kobayashi-Maskawa theory (CKM) has been shown to represent at least the dominant source of CP violation in $B_d$ and neutral $K$ 
transitions -- and that it is insignificant for the matter-antimatter asymmetry in our Universe. That asymmetry tells us 
New Dynamics (ND) with CP violation has to exist. The question is where can we find manifestation of such 
ND and at which level. 

The usual candidate for finding the dynamics underlying the Universe's baryonic asymmetry    
is neutrino oscillation with CP violation -- and we consider it a very good argument. However 
we find it wrong for our `roulette game' to place our `chips' on just one field. Our philosophy for probing CP 
invariance for ND is to cover all heavy flavour sectors -- strange, charm, beauty, top etc. -- and also search for EDMs. For charm decays and top productions and decays one has hardly any CP asymmetries from the Standard Model (SM). Dynamics of these quarks can therefore be well probed for the existence and analyses of ND without too much SM `background'. In charm transitions (unlike for top quarks) one has to deal with the impact of 
non-perturbative strong forces; while their impact creates 
non-trivial `homework', we want to show that the complexities caused by them offer awards too. 

The phenomenology of charm hadrons is richer than for strange ones. There are many more 
three- and four-body final states, and they cover much larger phase spaces. 
That means that branching ratios are smaller, expansions from chiral dynamics are larger and therefore less 
reliable than for kaons and hyperons. 

Furthermore long distance dynamics provide larger contributions, over which we have much less control, to decay rates than ND. 
Yet long distance dynamics cannot generate CP violation by themselves. 
Therefore we conclude that rare charm decays can hardly show clear manifestations of ND -- unless one 
observes CP violations, since the SM can produce only small asymmetries. 

The rich phenomenology of decays of charm hadrons represent great offers, not just vices: the much larger 
numbers of channels allow more tests of our description of the impact of hadronization. 
The treating of three- and four-body transitions requires much more `start-up' work; however,  
once this homework has been done satisfactorily, it  not only helps in understanding whether ND is behind the observed effects, but also about their features. 

For $D^{\pm}_{(s)}$ (and $\Lambda_c$ etc.) one can get only direct CP violation and, within the SM, 
only for singly Cabibbo suppressed non-leptonic decays -- except for final states with $K_S$ and $K_L$: 
for CP odd [even] components in $K_S$ [$K_L$] generate known CP asymmetry due to indirect CP violation 
in $K^0 - \bar K^0$ oscillation. ND could in principle generate direct CP violation in singly-  and 
doubly-Cabibbo suppressed (DCS) channels. The observed $D^0 - \bar D^0$ oscillations can generate indirect CP violation, and many classes of ND 
can do it well above the SM level.

In the following narrative, we will discuss CP violation in general for $D\to h_1h_2$, $h_1h_2h_3$ in Sect.\ref{HH} 
followed by the time evolution of the neutral $D$ eigenstates in Sect.\ref{DLDST}. Comments on the implication of assuming no direct CP violation on measured parameters are made in Sect.\ref{NODIRCPV} followed by a discussion on CP violation in $D^\pm_{(s)}$ in Sect.\ref{DSPM}. We look at 
$D^0\to h^+h^-l^+l^-$ in Sect.\ref{D0hhll}. Using B decays as charm factories is commented on in Sect.\ref{BFACT}. The correlation of final states in an $e^+e^-$ machine is discussed in Sect.\ref{EE}. then we describe 
$D_L \to h^+h^-l^+l^-$ transitions in relatively close analogy to 
$K \to \pi^+\pi^-e^+e^-$ in Sect.\ref{HHLL} and we comment on other four body final states with a lepton pairs in 
Sect.\ref{LLFINAL}. We discuss $D \to h l \nu$, $h\nu \bar \nu$ in Sect.\ref{HLNU} and comment on 
$D\to h_1h_2h_3h_4$ in Sect.\ref{HHHH}; then we consider the predictions from the class of Little Higgs Models 
with T parity (LHT) in Sect.\ref{LHMT} in detail and, in general, ND from non-minimal Higgs sectors in 
Sect.\ref{NONMINHIGGS} before concluding with a summary of our results in Sect.\ref{SUM}.

\section{\boldmath General Comments on CP Violations in $D\to h_1h_2$, $h_1h_2h_3$}
\label{HH}

\subsection{CPT Constraints}

CPT symmetry tells us 
the equalities of several classes of transitions; i.e., for singly- and doubly-Cabibbo suppressed one: 

\bea 
\Gamma (c\bar u \to q_1 \bar q_1 q_2\bar q_2)\; \;  &=& \; \; \Gamma (\bar c u \to \bar q_1 q_1 \bar q_2 q_2) \\
\Gamma (c\bar u \to d \bar s q\bar q)\; \;  &=& \; \; \Gamma (\bar c u \to \bar d s \bar q q) \\
\Gamma (c\bar d \to u \bar d q\bar q)\; \;  &=& \; \; \Gamma (\bar c d \to \bar u d \bar q q) \\
\Gamma (c\bar d \to u \bar s q\bar q)\; \;  &=& \; \; \Gamma (\bar c d \to \bar u s \bar q q)
\eea 
\bea 
\Gamma (c\bar s \to u \bar s q\bar q) \; \; &=& \; \; \Gamma (\bar c s \to \bar u s \bar q q) \\
\Gamma (c\bar s \to u\bar s d \bar s ) \; \; &=& \; \; \Gamma (\bar c s \to \bar u s \bar d s)
\eea 
with $q=u,d,s$. However, one has to measure transitions of hadrons: 
\bea 
\Gamma (D^+/D^0 \to [S=0])\; \;  &=& \; \; \Gamma (D^-/\bar D^0 \to  [S=0]) \\
\Gamma (D^+/D^0 \to [S=1])\; \;  &=& \; \; \Gamma (D^-/\bar D^0 \to [S=-1]) \\ 
\Gamma (D_s^+ \to [S=1]) \; \; &=& \; \; \Gamma (D^-_s \to [S=-1]) \\
\Gamma (D_s^+ \to [S=2] ) \; \; &=& \; \; \Gamma ( D_s^- \to [S=-2])
\eea 
Final states with $S=0$ are given with $\pi$'s, $\eta^{(\prime)} \pi$'s, $K\bar K\pi$'s, 
$S=1$ with $K^+\pi$'s, $K^+\eta, \ldots$ and $S=2$ with $K^+K^+\pi$'s and $K^+K^+\eta \pi$. 

Furthermore using strong FSI (ignoring SU(2) breaking in QCD and QED) these relations hold for final states separately depending on their isospin $I$. The equalities of {\em inclusive} rates allow asymmetries in 
{\em exclusive} rates, even sizable differences. These constraints show some practical usages in 
$D^+_{(s)} \to 2h$, $3h$, $4h$ decays, as we discuss below using also $G$ parity. 

For direct CP violation in rate $\Gamma (D\to f)$ vs. $\Gamma (\bar D\to \bar f)$ one needs two 
amplitudes with differences in {\em both} weak and strong phases. 
\begin{itemize}
\item
If the final states are produced by  
a single isospin amplitude, there can be no CP asymmetry -- no matter what weak dynamics can contribute whether it be from the SM or from ND. 
\item 
If two different isospin amplitudes can contribute, CP asymmetry can occur depending on SM and the 
impact of ND. In general one expects different landscapes for singly- and doubly-Cabibbo suppressed 
decays. For the leading source of indirect CP violation in beauty transitions one can use the Wolfenstein 
parametrization. However for CP asymmetries in charm decays one has to go to the CKM matrix through 
${\cal O}(\lambda ^6)$ and understand the differences between other parametrization as 
emphasized in Ref.\cite{HYCHENG}, where SM does not generate any CP asymmetries for 
doubly-Cabibbo suppressed decays 
and in singly-suppressed generates ones of order $\lambda^5$.  
\end{itemize}

\fancyhf{}

\lhead{\fancyplain{}{\nouppercase{\leftmark}}}
\cfoot{\fancyplain{}{\thepage}}
\boldmath 
\subsection{Lessons from LHCb Data on $D^0 \to K^+K^-/\pi^+\pi^-$}
\unboldmath 
In the decays $D^0 \to K^+K^-$ vs. $\bar D^0 \to K^+K^-$ and $D^0 \to \pi^+\pi^-$ vs. $\bar D^0 \to \pi^+\pi^-$, one 
can observe both direct and indirect CP violation. The very recent analysis presented by LHCb has searched for CP asymmetry in (mostly) time integrated 
rates in the {\em difference} of $D^0 \to K^+K^-$ and $D^0 \to \pi^+\pi^-$; its source is (mostly) 
direct $CP$ violation. They find \cite{LHCb1}:
\beq
\Delta A_{CP} \equiv A_{CP} (D^0 \to K^+K^-) - A_{CP} (D^0 \to \pi^+\pi^-) = 
- 0.82 \pm 0.21({\rm stat}.) \pm 0.11({\rm syst}.)
\eeq
with 3.5$\sigma$ deviation from zero. This is the first significant evidence for CP violation 
in $\Delta C \neq 0$ dynamics.  and it is important to infer whether it is due to SM or ND -- 
but has not been established experimentally yet. The next topic is more challenging, and just time will not solve it: 
Can we decide whether we observe the impact of ND? There one needs help from theorists -- and it will take time-consuming efforts in several ways. 
\begin{itemize}
\item
To get a realistic SM prediction for CP asymmetries in $D^0 \to 2h$ one has to understand the impact of 
final state interactions. Just diagrams showing them does not mean we can evaluate their impact. Those do not 
tell readers whether they are given by local or even short-distance dynamics or could reflect long-distance 
dynamics. 

Controlling the impact of final state interactions is not the strong point of lattice QCD studies. On the other hand 
two hadrons like pions or kaons are beyond the range of chiral dynamics. 
\item 
Looking at diagrams one can guestimate the scale for {\em inclusive} direct CP violation within SM of 
the order of $10^{-3}$. FSI can change CP asymmetries for exclusive channels significantly. Then one would want to 
probe three-, four-(and more) bodies for larger and smaller asymmetries. 
\item 
If direct CP violation in $D^0 \to K^+K^-$, $\pi^+\pi^-$ has been established -- and in particular on the level of several$\times 10^{-3}$ -- one thinks about local operators that can be the gate to ND. Such 
ND would affect CP asymmetries in $D^+ \to K^+K_S$ (not in $D^+ \to \pi^+\pi^0$ as explained below) and in 
$D^+_s \to K^+\phi$, $K^+\pi^0$, $K_S\pi^+$. 
\item 
It is important to study decays with three-body final states like $D \to K \bar K \pi$, $3 \pi$ 
and $D^+_s \to K^+K^+K^-$, $K^+\pi^+\pi^-$. Analyses of  
Dalitz plots take not only a lot of experimental work, but also theoretical efforts. The `Miranda' procedure \cite{MIRANDA} 
can be done in a model-independent way -- however one 
should {\em not} ignore theoretical help like from analyses of low-energy $hh$ scattering that also include 
experimental inputs through dispersion relations. The calculation effort for a single transition is 
sizable -- and can get help from symmetries arguments.  
\item 
One should analyze DCS decays like $D^0 \to K^+\pi^-$, $K^+\pi^-\pi^0$,  
$D^+ \to K^+\pi^0$, $K^+\pi^+\pi^-$ and $D^+_s \to K^+K_S$, $D^+_s \to K^+K^+\pi^-$. The rates for them are obviously lower, the SM `background' 
for direct CP violation is practically zero, since there is only a {\em single} SM amplitude.

\end{itemize}

For neutral charm mesons within CKM theory direct CP asymmetry has been predicted around 
${\cal O}(10^{-4})$ for $D^0 \to K^+K^-/\pi^+\pi^-$ with opposite signs 
(unless FSI affect them significantly)  and indirect CP violation, due to $|q/p| \neq 1$ and non-zero 
$\Im(\frac{q}{p}\bar A_f \otimes A_f)$, 
around ${\cal O}(10^{-3})$; the manifestation of the latter is suppressed by the CP insensitive observables $x_D$ 
and $y_D$, which are measured to be around 0.005 - 0.01. 
The experimental upper bounds on indirect CP violation in $D^0 \to h^+h^-$ are well above what the SM can produce. LHT, as a  {\em non-ad-hoc} class of ND, can produce values for indirect CP violation right up to those bounds, while hardly enhancing direct CP violation \cite{DtoHH}.  
For direct CP asymmetries in $D^0 \to K^+K^-/\pi^+\pi^-$ predictions have mainly been based on factorization for 
(quasi-)two-body final states; looking on diagrams gives only guestimates for the impact of FSI. {\em At present} 
one cannot prove that LHCb data are beyond SM dynamics. One has to decrease experimental uncertainty -- 
and the theoretical ones too; we will come back to it by emphasizing the importance of the analyses of three- and four-body 
final states. 

The sources of the asymmetry between $D^0 \to h^+h^-$ vs. $\bar D^0 \to h^+h^-$ can be differentiated by their 
dependence on the time of decay. To observe indirect CP violation one needs sensitivity to time of decays longer 
than $\tau_{D^0}$ -- as illustrated by the CDF analysis \cite{CDFDtoHH}. As ND like the LHT models can produce 
indirect CP violation right up to the experimental upper bounds; CDF data has reduced the range of their parameter space sizably. Alternatively, models with non-minimal Higgs sectors can enhance {\it both} direct and indirect CP violation in these channels which in turn makes experimental sensitivity to decay times much larger than $\tau_{D^0}$ even more important. The flavour-tagging of the initial neutral $D$ mesons can be done by the transition of $D^{*+} \to D^0 \pi^+$ vs. 
$D^{*-} \to \bar D^0 \pi^-$. This allows the measurement of the time of flight of the decaying neutral $D$ meson from its birth in general.

The transitions $D^0 \to K_S\phi$ and $\bar D^0 \to K_S\phi$ represent a promising place to exhibit 
a manifestation of ND, since the SM cannot generate direct CP violation (apart from $K_S \neq K_+$)\cite{DtoHH}. 
Yet a full analysis has to be performed of the Dalitz plot for $D^0/\bar D^0 \to K_S K^+K^-$; likewise 
for $D^0/\bar D^0 \to K_S \pi \pi$ with inclusion of $D^0/\bar D^0 \to K_S f_0(980)$ and 
$D^0/\bar D^0 \to K_S \sigma(600)$ contributions; while the total numbers of events in $D$ and 
$\bar D$ decays depend on their {\em productions}, the {\em relative} ratios in corresponding regions in both 
Dalitz plots {\em do not}. Moreover, direct vs. indirect CP asymmetries can be differentiated not only by their time evolution, but also their pattern in the 
locations in the Dalitz plots for three-body final states.

It is very unlikely 
that CP violation could produce $\Gamma (B^- \to D^0 D_s^-)\neq \Gamma (B^+ \to \bar D^0 D_s^+)$; 
therefore we can assume $\Gamma (B^- \to D^0 D_s^-) = \Gamma (B^+ \to \bar D^0 D_s^+)$; 
observation of $\Gamma (B^- \to h^+h^- D_s^-) \neq \Gamma (B^+ \to h^+h^- D_s^+)$ can be generated by 
CP violation in $D^0 \to h^+h^-$ vs. $\bar D^0 \to h^+h^-$. The same holds true for $B^- \to D^0 \pi^-$ vs. 
$B^+ \to \bar D^0 \pi^+$.

\section{\boldmath Time evolution of the neutral $D$ eigenstates}
\label{DLDST}

The decay rates of $D^0$ and $\bar{D}^0$ to their respective conjugate states are proportional to:
\begin{eqnarray}
\nonumber\left|T(D^0(t)\to f)\right|^2&=& \frac{1}{2}e^{-\bar{\Gamma}t}
\left[\left(\left|A_f\right|^2+\left |\frac{q}{p}  \right |^2\left|\bar A_f\right|^2\right)\cosh\left(y^{}_D\frac{t}{\tau_{D^0}}\right)
+\left(\left|A_f\right|^2-\left |\frac{q}{p}  \right |^2\left|\bar A_f\right|^2\right)
\cos\left(x^{}_D\frac{t}{\tau_{D^0}}\right)+ \right. \\
&&\left. 
+2\Re\left(\frac{q}{p}\bar A_f \otimes A_f^*\right)\sinh\left(y^{}_D\frac{t}{\tau_{D^0}}\right)-
2\Im\left(\frac{q}{p}\bar A_f \otimes A_f^*\right)
\sin\left(x^{}_D\frac{t}{\tau_{D^0}}\right)\right]\\
\nonumber\left|T(\bar{D}^0(t)\to \bar f)\right|^2&=& \frac{1}{2}e^{-\bar{\Gamma}t}
\nonumber \left[\left(\left|\bar{A}_{\bar f}\right|^2+\left|\frac{p}{q}\right|^2
\left| A_{\bar f}\right|^2\right)\cosh\left(y^{}_D\frac{t}{\tau_{D^0}}\right)+
\left(\left|\bar{A}_{\bar f}\right|^2-\left|\frac{p}{q}\right|^2
\left| A_{\bar f}\right|^2\right)\cos\left(x^{}_D\frac{t}{\tau_{D^0}}\right)+ \right. \\
&&
\left. +2\Re\left( \frac{p}{q}A_{\bar f}\otimes \bar A^*_{\bar f}\right)\sinh\left(y^{}_D\frac{t}{\tau_{D^0}}\right)-
2\Im\left( \frac{p}{q}A_{\bar f}\otimes \bar A^*_{\bar f}\right)\sin\left(x^{}_D\frac{t}{\tau_{D^0}}\right)\right]
\end{eqnarray}
It is assumed that $f$ and $\bar f$ are final states accessible to both $D^0$ and $\bar{D^0}$. Even the case of 
$D^0 \to l^-X^+$, $\bar D^0 \to l^+X^-$ is included, when $A_{D^0 \to l^-X^+} =0=\bar A_{\bar D^0 \to l^+X^-}$ holds. 
We use the symbol $\bar A_f \otimes A_f^*$ rather than just $\bar A_f A_f^*$: for {\em three}-body decays the 
total areas of Dalitz plots depend on the numbers of the produced charm mesons, but {\em not} the 
{\em relative} ratios between regions in the plot. 

In $D^0 \to \pi^+\pi^-$, $K^+K^-$, $K^+\pi^-$ vs. $\bar D^0 \to \pi^+\pi^-$, $K^+K^-$, $K^+\pi^-$ the differences 
between direct and indirect CP violations can be shown only due to their time evolutions of the partial widths. 
For non-leptonic $D^0 \to h_1h_2h_3$ vs. $\bar D^0 \to \bar h_1 \bar h_2 \bar h_3$ it was mentioned that 
the pattern of CP asymmetries can be differentiated by their locations in the Dalitz plots.

Allowing for {\it both} direct and indirect CP violation, presently available data lead to \cite{HFAGCHARM}\footnote{Up to date results can be found in the \href{http://www.slac.stanford.edu/xorg/hfag/charm/CHARM10/results_mix+cpv.html}{HFAG} website}: 
\bea
\nonumber x^{}_D = \frac{\Delta M_D}{\Gamma_D} = \left( 0.63 ^{+0.19}_{-0.20}\right) \% \; &,& \; 
y^{}_D = \frac{\Delta \Gamma _D}{2\Gamma_D} = \left( 0.75 \pm 0.12\right) \% \\
\left |\frac{q}{p}  \right | = 0.89 ^{+0.17}_{-0.15} \; &,& \;  
\phi^{}_D   = \left( - 10.1 ^{+9.4}_{-8.8}\right) ^o
\label{DOSCDATA}
\eea
-- i.e., a very different pattern from neutral kaons:  
\begin{itemize}
\item 
The `mass' eigenstates are close to flavour ones and far from being CP eigenstates, and cuts on lifetimes hardly produce the latter. 
\item 
While the experimental value of $|q/p|$ for $D^0$ meson is within one half sigma of the benchmark 
unity for no CP violation, it would also be within one sigma of a value of 0.7 -- i.e., large indirect CP asymmetry. 
\item 
ND can produce such large indirect CP violation -- and can generate sizable direct CP asymmetries in 
once and twice Cabibbo suppressed transitions, although not necessarily both can be produced by the same class of ND.

\item
Therefore, one needs excellent time resolution to differentiate indirect and direct CP violation, 
and one needs to compare time evolutions of $D^0$ and $\bar D^0$ transitions. Realistically 
one can retain only terms through first orders in $x^{}_D$ and $y^{}_D$. 
\begin{itemize}
\item 
For $D^0 \to l^-X$ vs. $\bar D^0 \to l^+X$ we can use:
\bea 
|T(D^0(t) \to l^-X)|^2 \propto e^{-\bar \Gamma t} \left| \frac{q}{p}\right|^2 \left|A_{\bar D^0 \to l^-X}  \right|^2 \\
|T(\bar D^0(t) \to l^+X)|^2 \propto e^{-\bar \Gamma t} \left| \frac{p}{q}\right|^2 \left|A_{D^0 \to l^+X}  \right|^2
\eea
with $\left|A_{\bar D^0 \to l^-X}  \right| = \left|A_{D^0 \to l^+X}  \right|$. The indirect CP asymmetry  between 
the `wrong-charge' semileptonic rates are time independent: 
\beq
\frac{\Gamma (D^0 \to l^-X) - \Gamma (\bar D^0 \to l^+X)}{\Gamma (D^0 \to l^-X) + \Gamma (\bar D^0 \to l^+X)}
= \frac{|q/p|^2 -|p/q|^2 }{|q/p|^2 + |p/q|^2}
\eeq
To find the time integrated rates one has to include second orders in $x^{}_D$ and $y^{}_D$ to get 
\bea
\Gamma (D^0 \to l^-X) \propto \frac{x^2_D + y^2_D}{2}\left| \frac{q}{p} \right|^2 \\
\Gamma (\bar D^0 \to l^+X) \propto \frac{x^2_D + y^2_D}{2}\left| \frac{p}{q} \right|^2
\eea
While those rates are tiny due to the size of $x^2_D + y^2_D$ -- yet the CP asymmetry might be sizable due to 
$|q/p| \neq 1$. 

\item 
$D^0$, $\bar D^0 \to K_S \phi$ come mostly from Cabibbo favoured transition, where contributions from ND and direct CP violation 
is very unlikely to occur. Therefore we use use $|A_{D^0 \to K_S\phi}|= |A_{\bar D^0 \to K_S\phi}|$ and get the 
simple expressions: 
\bea 
|T(D^0(t) \to K_S\phi)|^2 \propto e^{-\bar \Gamma t}\left[ 1 + \frac{t}{\tau _D}\left| \frac{q}{p}\right| 
\left( y^{}_D \cos\phi^{}_D - x^{}_D \sin\phi^{}_D   \right)      \right] \\
|T(\bar D^0(t) \to K_S\phi)|^2 \propto e^{-\bar \Gamma t}\left[ 1 + \frac{t}{\tau _D}\left| \frac{p}{q}\right| 
\left( y^{}_D \cos\phi^{}_D - x^{}_D \sin\phi^{}_D   \right)      \right] 
\eea
For indirect CP violation one thus gets
\beq
\frac{|T(D^0(t) \to K_S\phi)|^2 - |T(\bar D^0(t) \to K_S\phi)|^2}
{|T(D^0(t) \to K_S\phi)|^2 + |T(\bar D^0(t) \to K_S\phi)|^2} \simeq 
\frac{1}{2}\frac{t}{\tau _D}\left[ y^{}_D \left(\left| \frac{q}{p}\right| - \left| \frac{p}{q} \right| \right) \cos\phi^{}_D
- x^{}_D  \left(\left| \frac{q}{p}\right| + \left| \frac{p}{q} \right| \right)\sin\phi^{}_D \right] \; , 
\eeq
i.e., another sign of indirect CP violation that also includes $\phi^{}_D \neq 0$: 
\begin{itemize}
\item 
The term proportional to $y^{}_D$ is driven by $|q|\neq |p|$, yet decreased by $\phi^{}_D \neq 0$; the term proportional to $x^{}_D$ is driven 
by $\phi^{}_D \neq 0$ and enhanced by $|q| \neq |p|$. 

\end{itemize}
\item 
Once Cabibbo suppressed decays like $D \to K^+K^-$ and $D\to \pi^+\pi^-$ gets both indirect and 
direct CP violation already in the CKM theory, however these are very small asymmetries: 
$|A|^2/|\bar A|^2$ differs from unity by $10^{-4}$ or less, $\left| 1 - |q/p|\right|$ and 
$\phi^{}_D$ are generated to about $10^{-3}$. A class of non-ad-hoc ND like LHT can generate $\left| 1 - |q/p|\right|$ 
up to even 0.3 and $|\phi^{}_D|$ up to 30$^\circ$ -- i.e. up to present experimental bounds. However, it is possible, but harder 
to obtain direct CP violation in these decays much larger than what comes from CKM theory. On the other hand, CP violating phases in models with non-minimal Higgs sectors can enhance direct CP violation in these channel.
\item 
For DCS decays like $D^0 \to K^+\pi^-$ vs. $\bar D^0 \to K^-\pi^+$  
time dependent CP asymmetry should be much larger due to the interference with $\Delta C=1$ 
dynamics through $A_{\bar D^0 \to K^+\pi^-}/A_{D^0 \to K^+\pi^-}$ and 
$A_{D^0 \to K^-\pi^+}/A_{\bar D^0 \to K^-\pi^+}$. Finally in the SM one gets 
$$|A_{D^0 \to K^+\pi^-}|^2 = |A_{\bar D^0 \to K^-\pi^+}|^2 \;\;{\rm and}\;\; 
|A_{D^0 \to K^-\pi^+}|^2 = |A_{\bar D^0 \to K^+\pi^-}|^2$$ as explained in the beginning.  
For $c\to d \bar s u$ transitions one gets a combination of $\Delta I = 0\,  \& \, 1$ amplitudes 
-- i.e., two different strong phases; ND like the exchange of charged Higgs states can generate 
a weak phase different from the CKM theory. Therefore one can get also a {\em direct} CP asymmetry due to
$|A_{D^0 \to K^+\pi^-}|^2 \neq |A_{\bar D^0 \to K^-\pi^+}|^2$.

\end{itemize}
\end{itemize}
Resume for $D^0$ vs. $\bar D^0$ transitions: 
\begin{enumerate} 
\item 
There are three classes of observables sensitive to CP violation: 
(i) $|q/p| \neq 1$, which is purely indirect CP violation. (ii) $|A_f| \neq |\bar A_{\bar f}|$ is due to 
direct CP violation and depends on the final states. (iii) The phase between $q/p$ and 
$\bar A_f/A_f$ due to the interference between $\Delta C=2$ and $\Delta C=1$ dynamics; 
direct CP violation will in general induce a different phase for two different final states. 
\item 
One can study CP asymmetries in $D^{*+} \to D^0 \pi^+ \to [l^-X]\pi^+$ vs. $D^{*-} \to \bar D^0 \pi^- \to [l^+X]\pi^-$ 
and $D^{*+} \to D^0 \pi^+ \to [K_S\phi/K^+K^-/\pi^+\pi^-/K^+\pi^-]\pi^+$ vs. 
$D^{*-} \to \bar D^0 \pi^- \to [K_S\phi/K^+K^-/\pi^+\pi^-/K^-\pi^+]\pi^-$. Such comparisons give us 
more validations for the measurements and more information on the direct and indirect CP violations 
and the features of the underlying ND.  As mentioned above one can analyze $B^- \to D^0 D_s^-$ vs. 
$B^+ \to \bar D^0 D_s^+$. 
\item 
Searches for time dependent CP asymmetries have to face the challenge of dealing with $y^{}_D$ and $x^{}_D$ of 
1 \% or less. Furthermore for small effects around ${\cal O}(10^{-4} - 10^{-3})$ one has to worry about 
flavour-tagging. 
\end{enumerate}

\section{Implications of assuming {\bf{\em no direct}} CP violation in charm dynamics.}
\label{NODIRCPV}

In a previous work \cite{DtoHH} we had suggested that the $D^0 \to \pi^+\pi^-/K^+K^-$ data \cite{CDFDtoHH} should be interpreted as being driven by indirect CP violation and not by direct CP violation if the ND involved belongs to the LHT-like framework. The Heavy Flavour Averaging Group (HFAG) \cite{HFAGCHARM} has recently made an analysis of the oscillation and CP violation parameters assuming that there is no direct CP violation in charm dynamics. Only two of the listed parameters suffered additional constraints under such an assumption, namely, $|q/p|$ and $\phi^{}_D$. The quoted values are:
\begin{equation}
\left |\frac{q}{p}  \right | = 1.02 \pm 0.04\;,\;  
\phi^{}_D   = \left( - 1.05 ^{+1.89}_{-1.94}\right) ^o
\label{eq:NODCPV}
\end{equation}

Throughout this article we have looked at two classes of ND which have two different contributions to CP violation.
\begin{itemize}
\item One class belongs to the LHT-like framework and can enhance indirect CP violation in charm by orders of magnitude above the SM estimate. However, this class of model cannot enhance direct CP violation and hence all sources of direct CP violation has to come mostly from the SM.
\item The other class belongs to models with extra Higgs sectors and can enhance both direct and indirect CP violation over and above what the SM can generate. 
\end{itemize}

Of course, there are always possibilities that ND manifests itself as a hybrid of these two classes of models. The numbers quoted in eq.\ref{eq:NODCPV} applies to model that belong to the LHT-like framework and we shall use these constraints when discussing the effects of ND of this class. However, an analysis of how the limits on $|q/p|$ and $\phi^{}_D$ constrain the parameter space of LHT-like models will be left for a future work. 

\section{Direct CP Asymmetries in $D_{(s)}^{\pm}$ Decays}
\label{DSPM}

No matter if direct CP asymmetry in $D^0 \to K^+K^-/\pi^+\pi^-$ is established or not, there are strong reasons 
for the existence of ND and the study of its shape -- in particular in DCS $D^{\pm}_{(s)}$ decays.

\subsection{Singly-Cabibbo Suppressed Decays}

Ignoring isospin breaking in QCD and corrections from QED there can be {\em no} CP asymmetry in 
$D^{\pm} \to \pi^{\pm}\pi^0$ even with ND, since there is only one $I=2$ amplitude and therefore 
no strong phase difference.

The same 
holds for $D^+\to K^+K_S$ vs. $D^- \to K^-K_S$ (apart from $K_S$ not being a pure CP eigenstate): 
$c\to s \bar s u$ represents a $\Delta I = 1/2$, $\Delta I_3 = + 1/2$ transition and therefore cannot mix with 
$D^+ \to \pi^+\pi^0$ in the limit of isospin symmetry. 

For $D^+ \to K_S\pi^+$ vs. $D^- \to K_S \pi^-$ one can (even in addition to $K_S\neq K_+$) have a 
difference due to ND affecting the DCS transition $c \to d \bar s u$ that interferes with the SM amplitude $c \to s \bar d u$ for $D^+ \to K_S\pi^+$. Due to the known CP asymmetry in the neutral K system this asymmetry can be predicted to be: 
\begin{equation}
\frac{\Gamma(D^+\to K_S\pi^+)-\Gamma(D^-\to K_S\pi^-)}{\Gamma(D^+\to K_S\pi^+)+\Gamma(D^-\to K_S\pi^-)}\simeq2{\rm Re}(\epsilon_K)\simeq 3.3\times10^{-3}
\end{equation}
If ND contributes significantly to the DCS amplitude $c \to d \bar s u$, it will probably change 
this  prediction; sizable FSI between $K^0\pi^+$ and $\bar K^0 \pi^+$ are expected.  

In analogy to $D^{\pm}$ decays one predicts a CP asymmetry in $D^+_s \to K_SK^+$ vs. $D^-_s \to K_S K^-$ 
of $3.3\times10^{-3}$; an impact of ND on the amplitude for $c\to d \bar s u$ can affect this prediction. 
For Cabibbo suppressed transitions $D^+_s \to K^+\pi^0$ vs. $D^-_s \to K^- \pi^0$ one predicts asymmetry of order 
$10^{-5}$ to $10^{-4}$ in the  CKM theory; for $D^+_s \to \pi^+K_S$ vs. $D^-_s \to \pi^-K_S$ one expects asymmetry up to 
$4.3\times10^{-3}$ due to direct CP violation in $\Delta C=1$ CKM dynamics.

There are subtle features here: 
\begin{itemize}
\item 
Through order $\lambda ^6$ there is no {\em weak} phase for these DCS transitions within the SM; therefore there can be no CP asymmetry on the quark level, never mind FSI. 
\item 
There are weak phases of order $\lambda ^4$ for Cabibbo favoured transitions. However the same weak phase 
controls the amplitudes, therefore no CP asymmetry can be produced at the quark level. 

\item 
Yet one measures transitions at the hadronic level: $D^+ \to K^+h's$ and 
$D^+_s \to K^+K^+ h's $ (with $h$ denoting $\pi's$ and $\eta/\eta^{\prime}$), but also 
$D^+ \to K_{S,L} +h's$ and 
$D^+_s \to K^+K_{S,L} +h's/K_{S,L}K_{S,L}+h's $. Those mix Cabibbo favoured and doubly suppressed amplitudes. 

\end{itemize}

\subsection{Doubly-Cabibbo Suppressed Decays}

In $D^{\pm} \to K^{\pm}\pi^0$ decays the final states contains interference between $I=1/2$ and $I=3/2$ 
amplitudes with different strong phases. In the SM there are no weak phases through order $\lambda^6$ 
-- i.e., no practical CP asymmetry. However ND can generate sizable weak phase to compete with doubly-Cabibbo 
suppressed amplitudes. The landscape for $D^{\pm} \to K^{\pm}\pi^+\pi^-/K^{\pm}K^+K^-$ is even richer, 
as discussed below. 

For $D_s$ decays one can measure $D_s^{\pm} \to K^{\pm}K_s$ -- but asymmetries can `hide' in 
Cabibbo favoured decays. However one can probe $D_s^+ \to K^+K^+\pi^-$ 
vs. $D_s^- \to K^-K^-\pi^+$.

\boldmath
\subsection{$D^{\pm}_{(s)} \to h_1h_2h_3$}
\unboldmath

The decays of charged $D$ mesons to final states with three pseudo-scalars give us much more information about 
the dynamics underlying direct CP violation than studying final states with two pseudo-scalars, where one 
can measure only differences in the partial widths. For three-body final states one can measure -- and it has been done -- the two-dimensional Dalitz plot with the known resonances. There one can differentiate strong vs. weak phases and provide more information about the possible impact of and the nature of ND. 
While performing a full Dalitz plot analysis requires more data and time -- even using the `Miranda' procedure 
\cite{MIRANDA} -- it gives us a great deal of information on the ND; in particular one has to include the informations about contributions from scalar-pseudoscalar final states like $K\sigma$, $K\kappa$ and $\pi\sigma$. 
{\em Most CP asymmetries do not depend on the relative $D_{(s)}^+$ vs. $D_{(s)}^-$ productions.} 
There are three classes of CP asymmetries:
\begin{enumerate}[I.]
\item 
A difference in $D \to Rh$ vs. $\bar D \to \bar R \bar h$; 
\item 
Difference in the interference between $D \to R_1 h_1$ and $D \to R_2 h_2$ vs. 
$\bar D \to \bar R_1 \bar h_1$ and $\bar D \to \bar R_2 \bar h_2$; 
\item 
Difference in true three-body final states or broad resonances.  
\end{enumerate}
For classes I and II one can use Breit-Wigner prescription for resonances even for an accurate analyses; however 
for class III one needs some input from theory to use dispersion relations from $h_1h_2$ scattering at low energy. 
Several studies are at work. 

In our previous paper \cite{DtoHH} we have shown that LHT models for ND do not have a good chance to 
manifest themselves through direct CP violations in charm decays. However ND models with a charged Higgs sector could have a sizable impact, in particular for final states with scalar resonances like $Kf_0(980)$, 
$K\sigma (600)$, $\pi \sigma (600)$ etc.


\section{\boldmath $D^0 \to h^+h^-l^+l^-$ vs. $\bar D^0 \to h^+h^-l^+l^-$}
\label{D0hhll}

In $K \to \pi^+\pi^- \gamma$ one has partial widths and the T {\em odd} correlation 
$\langle \vec \epsilon _{\gamma} \cdot (\vec p_+ \times \vec p_- ) \rangle$, where $\vec \epsilon _{\gamma}$ 
describing the photon polarization. The most realistic way to infer $\vec \epsilon _{\gamma}$ is based on 
$K^0/\bar K^0 \to \pi^+\pi^- \gamma^* \to \pi^+\pi^- e^+e^-$ and in measuring the angle $\Phi$ between the 
$\pi^+ - \pi^-$ and $e^+ - e^-$ planes. 

The largest CP asymmetry in kaon decays has been found in $K_L \to e^+e^-\pi^+\pi^-$, namely this 
T odd correlation, a prediction by 
Sehgal and Wanninger \cite{SW}
\beq
\left. A_{\rm T}\right|_{\rm theory} = (14.3 \pm 1.3) \% 
\eeq
that was measured later by the KTeV and NA48 experiments  
\beq
\left. A_{\rm T}\right|_{exp} = (13.7 \pm 1.5) \% \; . 
\eeq
The large effect is generated by the small quantity $|\eta _{+-}|$ in the $K_L$ wave function. 
As usual a price has to be paid for such an enhancement, namely a tiny branching ratio of 
$(3.11 \pm 0.19)\times 10^{-7}$. In the context of beauty physics, such kinematically nontrivial CP violation stemming from triple products of spin and momenta was discussed in \cite{BK}. This can bring about some hope of finding an enhancement in 
$D \to l^+l^-h^+h^-$ too with $l=e$, $\mu$ and $h=K$, $\pi$ again with the normal price of a tiny branching fraction.

It should be noted that the analogy between such $D^0$ with $K^0$ decays is only qualitative: 
\begin{itemize}
\item 
The lifetimes of the two mass eigenstates in the $K^0 - \bar K^0$ system are very different so one can separate 
$K_L$ and $K_S$ easily; therefore the notation is obvious. The lifetime difference in the $D^0 - \bar D^0$ complex is only 2\% or less and hence  one has to compare $D^0$ and $\bar D^0$ time evolutions carefully. To identify a T odd correlation as a CP asymmetry one has to compare them in $D^0$ and $\bar D^0$ 
transitions and show that they are {\em not} of opposite signs. 
\item 
The mass eigenstates $K_L$ and $K_S$ are almost aligned with the {\em odd} and {\em even} CP eigenstates. For the 
$D^0 - \bar D^0$ complex both mass eigenstates might contain sizable 
odd and even components in the presence of ND. 
\item 
Unlike for the neutral kaon system, direct CP asymmetry might not be smaller than the indirect CP effects for the neutral $D$ mesons. Like in the past, we shall continue to assume that the contribution from one class of ND, namely LHT-like ND, to direct CP violating dynamics is tiny compared to the indirect one. If reality proves to be in the contrary, it will only serve to enhance the effects that we speak of while a quite different classes of ND may have to be considered when models like the ones with non-minimal Higgs sectors might be good candidates.

\item 
While the indirect CP violation was known from $K_L \to \pi^+\pi^-$, we have only an upper bound for $D^0$ 
transitions. Furthermore it is (and was) known that CP asymmetries in neutral kaons is dominated by 
indirect CP violation. However we do not know what pattern holds for indirect vs. direct CP violations in 
neutral charm transitions. 

\item 
$K$ decays can be treated in chiral dynamics with `soft' pions  but this can hardly be trusted in $D$ decays 
to pions, kaons and photons. In any case one cannot expects similar effects. 
\item
At the same time we can see also virtues, not just vices, namely to have 
$D \to K\pi l^+l^- , K \bar K l^+l^-  , \pi \pi l^+ l^-$ with $l = e , \mu$ and 
$D \to \pi \pi \gamma , K \bar K \gamma , K\pi \gamma$ for comparisons of M1, E1 and IB amplitudes 
to check predictions. 

\end{itemize}
In terms of the flavour eigenstates the decay can be expressed as:
\begin{eqnarray}
\frac{d}{d\Phi}\Gamma(D^0(t)\to h^+h^-l^+l^-)&=& \Gamma^0_1(t)\; {\cos}^2\Phi + \Gamma^0_2(t)\;{\sin}^2\Phi + 
  \Gamma^0_3(t)\; {\cos}\Phi \;{\sin}\Phi\\
 \frac{d}{d\Phi} \Gamma(\bar{D}^0(t)\to h^+h^-l^+l^-)&=& \bar{\Gamma}^0_1(t)\; {\cos}^2\Phi + \bar{\Gamma}^0_2(t)\;{\sin}^2\Phi - 
  \bar{\Gamma}^0_3(t)\; {\cos}\Phi \;{\sin}\Phi \; .
\end{eqnarray}
The change in sign of $\Gamma_3$ vs. $\bar \Gamma_3$ is  due to $\Phi\to -\Phi$ under a CP transformation 
as CP symmetry implies $\Gamma_{1,2} = \bar \Gamma_{1,2}$ and 
$\Gamma_3 = - \bar \Gamma_3$. 

CP violation, given by $\Gamma_{1,2} \neq \bar \Gamma_{1,2}$ can be driven by direct CP violating dynamics only. However, $\Gamma_3\neq\bar{\Gamma}_3$, can be driven by both direct and indirect CP violating dynamics. The asymmetry can be written as:
\begin{eqnarray}
\frac{\Gamma(D^0(t)\to h^+h^-l^+l^-)-\Gamma(\bar{D}^0(t)\to h^+h^-l^+l^-)}{\Gamma(D^0(t)\to h^+h^-l^+l^-)+\Gamma(\bar{D}^0(t)\to h^+h^-l^+l^-)}=a^{dir}_{\rm CP}+a^{ind}_{\rm CP}\frac{t}{\tau_{D^0}}
\label{eq:assymhhll}
\end{eqnarray}
where $a^{dir}_{\rm CP}$ is the time independent part which is driven by direct CP violation and $a^{ind}_{CP}$ is the time dependent part which is primarily driven by $D^0 - \bar{D}^0$ oscillations but can have contributions from direct CP violating dynamics too. In the absence of sizable direct CP asymmetry in charm dynamics, i.e. $|T(D^0\to h^+h^-l^+l^-)|=|T(\bar{D}^0\to h^+h^-l^+l^-)|$,  it can be safely assumed that the entire asymmetry depicted in Eq.\ref{eq:assymhhll} is driven by oscillations. However, there is some subtlety that needs to be considered here. The time dependent term $a^{ind}_{\rm CP}$ is given by
\begin{eqnarray}
a^{ind}_{\rm CP}=\frac{1}{2}\left[y^{}_D\Re(\lambda-\lambda^{-1})-x^{}_D\Im(\lambda-\lambda^{-1})\right] \; , \; 
\lambda=\frac{q}{p}\frac{T(D^0\to h^+h^-l^+l^-)}{T(\bar{D}^0\to h^+h^-l^+l^-)}
\end{eqnarray}
In the absence of direct CP violations we have 
\begin{equation}
\lambda=\left|\frac{q}{p}\right|e^{i(\phi+\delta)}
\end{equation}
\begin{figure}[b!]
\includegraphics[width=17cm]{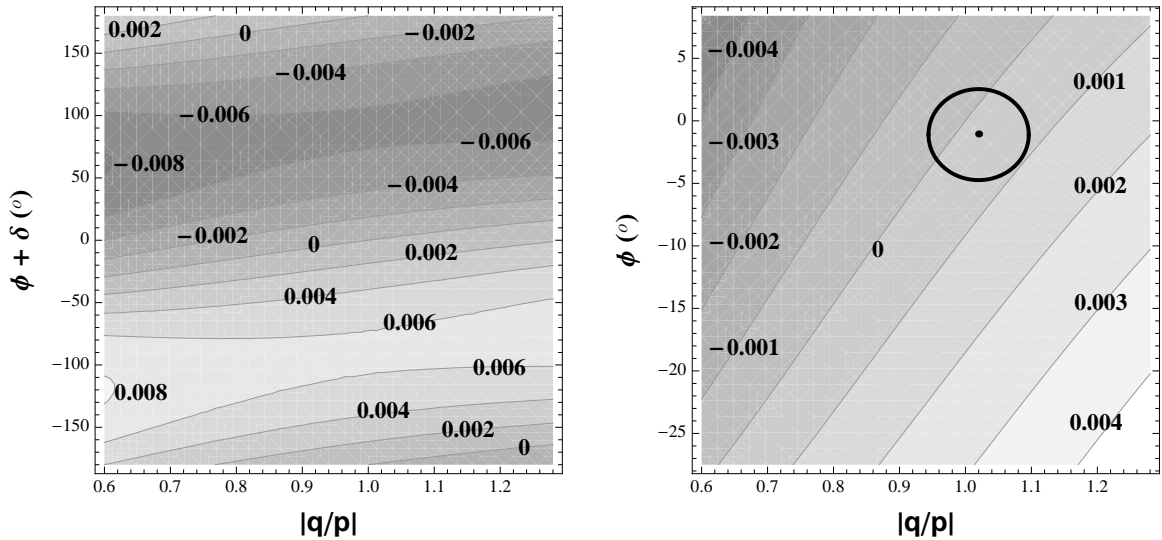}
\caption{Contour plot of the indirect CP asymmetry parameter $a^{ind}_{\rm CP}$. The plot on the left allows for any arbitrary value of $\delta$ in the range $\{-\pi,\pi\}$. The ellipse marks the possible asymmetry assuming no direct CP violation with the dot marking the central values under such an assumption.  The $|q/p|$ and $\phi$ values used here are the $95\%$ CL limits as quoted in \cite{HFAGCHARM}. }
\label{fig:indirqpphi}
\end{figure}
\begin{figure}[h!]
\includegraphics[width=17cm]{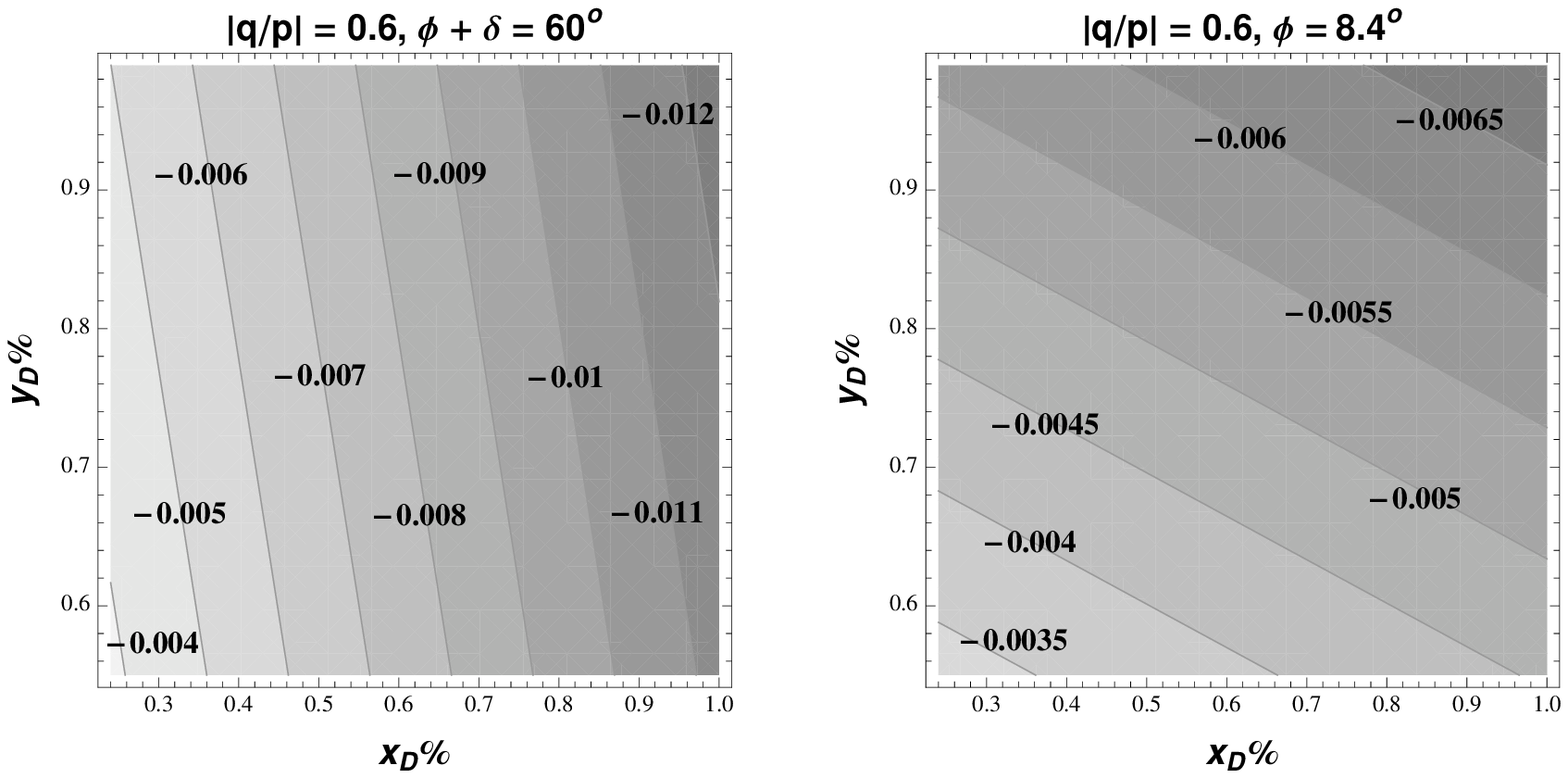}
\caption{Contour plot of the indirect CP asymmetry parameter $a^{ind}_{\rm CP}$. The plot on the left allows for a finite value of $\delta$. The $|q/p|$ and $\phi$ values used here are the $95\%$ CL limits as quoted in \cite{HFAGCHARM}.}
\label{fig:indirxy}
\end{figure}
The angle $\phi$ is the phase of $q/p$ and comes from mixing of the flavour eigenstates and hence contributes to oscillations. However, the phase $\delta$ is final state dependent as it comes from final state interactions (FSI) and is the phase difference between  $T(D^0\to h^+h^-l^+l^-)$ and $T(\bar{D}^0\to h^+h^-l^+l^-)$ which can be present even in the absence of direct CP violation in this decay channel. This argument holds true for any final state if oscillations can drive a CP asymmetry in it. Hence a genuine difference in $a^{ind}_{\rm CP}$ between different decay modes can be an indirect signal of direct CP violation.

While the observation of $D^0-\bar D^0$ oscillations seems established, the relative size of $x_D$ and 
 $y_D$ is not clear yet. {\em Before} these experimental results in 2007, most authors had argued that the 
 SM predicts $x_D$, $y_D$ $\leq 3\times 10^{-4}$ --- yet not all. In 1998, 
 $x_D$, $y_D$ $\leq 10^{-2}$ was listed, admittedly as a {\em conservative} SM bound \cite{VAR98}, 
 together with a question: How can one rule out that the SM can not produce 
 $10^{-6} \leq r_D \leq 10^{-4}$ (corresponding to $x_D$, $y_D$ $\sim 10^{-3} - 10^{-2}$). 
 In 2000 and 2003, an SM prediction obtained from an operator product expansion (OPE) yielded $x_D$, $y_D$ $\sim {\cal O}(10^{-3})$ 
 \cite{DUAL};  later a more sophisticated OPE analysis was done 
with similar results \cite{LENZ}. Alternatively 
 in 2001 and 2004, an SM prediction on $D^0-\bar D^0$ oscillations was based on $SU(3)$ breaking mostly in the phase space for $y_D$ and then from a dispersion relation for $x_D$ \cite{FALK}.
 
Even though the present experimental results on $x_D$ and $y_D$ can be accommodated within some available theoretical 
SM estimates, ND can play a significant role here  \cite{BLUM,DMB,DKdual}. In particular in \cite{DKdual} such has been 
analyzed for LHT-like dynamics, if future data confirm that $x_D$ indeed falls in the range of $0.5$\% and $1$\%. While SM 
long distance dynamics could accommodate a value in that range, it could not be ruled out that ND could contribute 
half or quarter of the value of $x^{}_D$ considering the reasonable theoretical uncertainty in the predictions from 
long distance dynamics. Yet indirect CP violation in $D^0 - \bar D^0$ oscillations would provide us a clear signature 
for the manifestation of ND.

From the theoretical SM estimates for $x_D$ and $y_D$, which are hardly more than guesses since the SM predictions depend strongly on our 
theoretical treatment of LD dynamics for $x_D$ and $y_D$ (or its lack thereof) and our treatment 
of the extraction of the CKM phase as it enters charm decays \cite{DUAL,DMB} one gets a SM ``prediction'' of:
\beq
a^{\rm ind}_{\rm CP}(t)|_{\rm SM} \sim O(10^{-5})  \; , 
\eeq
while the same for direct CP violation is most likely
\beq
a^{\rm dir}_{\rm CP}(t)|_{\rm SM} \sim O(10^{-4})
\eeq
for the final states where $h^+h^-=\pi^+\pi^-/K^+K^-$. If we consider $D^0 \to K^+\pi^- l^+l^-$ vs. 
$\bar D^0 \to K^-\pi^+ l^+l^-$, SM cannot generate direct CP violation in this DCS channel. The contribution to the decay rates from the SM are smaller too and hence there can be possible large ND intervention here.

In Fig.\ref{fig:indirqpphi} we show how large $a^{ind}_{\rm CP}$ can be for the current central value of $x^{}_D$ and $y^{}_D$. We have varied $q/p$ and $\phi$ over the current $95\%$ CL values. For the figure on the left we have allowed an arbitrary variation of $\delta$ over the entire range of $\{-\pi,\pi\}$. In Fig.\ref{fig:indirxy} the same is depicted but over variations of $x_D$ and $y_D$ over their current $95\%$ CL values with a choice of $q/p$ and $\phi$ where the $a^{ind}_{\rm CP}$ is maximized for these parameters. It can be clearly seen that at most
\begin{equation}
a^{ind}_{\rm CP}\simeq10^{-2}
\end{equation}
Such time dependent CP asymmetry can be measured in hadron machines as has been done for 
$D^0\to\pi^+\pi^-,K^+K^-$ at CDF recently \cite{CDFDtoHH}. What is important to distinguish between indirect and direct CP violations is that experimental analysis of such time evolution should be conducted over large values of $\langle t \rangle/\tau_{D^0}$. Possibilities of measuring the asymmetries at varying values of $\langle t\rangle/\tau_{D^0}$ ranging from $\langle t\rangle/\tau_{D^0}=1$ to $\langle t\rangle/\tau_{D^0}\gg 1$ should be looked into. Later we will show that ND can come close to such upper bound, namely from  LHT-like models or models with 
non-minimal Higgs sectors. However, assuming no sources of direct CP violation in charm dynamics puts additional constraints on models within the LHT-like framework. 

Akin to the analysis done for $K^0\to\pi^+\pi^-e^+e^-$ in \cite{SvL} a ``time-resolved'' analysis can be done for the $D^0$ system too. The charm specific analysis would display the evolution of the asymmetry from its zero value at short time scales to its maximum value at large time scales. A spectrum integrated asymmetry would display the interference between the two mass eigenstates and an asymmetry between the decay rates of $D^0$ vs. $\bar{D}^0$ can also be analyzed. At this time such a study is not feasible as we have very little knowledge of the strong phases involved in the decays $D^0\to h^+h^-$. However, such an analysis would be quite instrumental in determining what scales of $\langle t\rangle/\tau_{D^0}$ to probe for experimentally. It should be kept in mind that due to the tiny lifetime difference between the two mass eigenstates of the $D^0$ meson, time resolution of the order of the difference might be necessary to do a time resolved study, a capability that shall, hopefully, exist in the future. 

The plots clearly show that the observable CP asymmetries cannot be generated above few$\times 10^{-3}$ in general. That is not surprising, 
since the CP insensitive parameters $x_D$ and $y_D$ -- where ever they come from -- are around 0.005 - 0.01: 
ND cannot generate more than that range. 

\section{\boldmath $B^{\pm}$ Decays as a Factory of Neutral $D$ Mesons}
\label{BFACT}

The decays of charged $B$ mesons can act as a `filtered' factory for neutral $D$ mesons with sizable 
branching ratios: 
\bea
{\rm BR}(B^+ \to \bar D^0 D_s^+) &=& 0.010 \pm 0.0017
\\ 
{\rm BR}(B^+ \to \bar D^0 D_s^{*+}) &=& 0.0076 \pm 0.0016
\eea
Searches for small CP asymmetries in neutral $D$ transitions in 
three- and four-body final states should be possible at future Super-B factories, where 
{\em at least partially} time integrated asymmetries in final state distributions should be possible.

\boldmath
\section{\boldmath On $e^+e^- \to D_1 D_2$}
\label{EE}
\unboldmath

If one would have correlated $D^0 - \bar D^0$ pairs, one could 
find larger effects. It can happen in $e^+ - e^-$ annihilation realistically at the $\psi''(3770)$ resonance. 

Possible CP asymmetries in $D$ transitions are very small and the distinction of indirect vs. direct 
CP violation over time evolutions very subtle.  Then the question arises that quantum correlations in 
$e^+e^- \to D^0 \bar D^0 \to D_S D_L$ might help us here in some qualitative analogy with 
$e^+e^- \to \phi (1020) \to K^0 \bar K^0$: there one can filter the mass eigenstates using 
$e^+e^- \to K_SK_L \to [\pi^+\pi^-]_{K_S}K_L$ 
as a flavour-tagged production of $K_L \to 2\pi$, $e^+e^-\pi^+\pi^-$ due to $\tau _{K_S} \ll \tau _{K_L}$. 
For 
\beq
e^+e^- \to \psi''(3770) \to D^0\bar D^0/D_+D_-/D_1D_2 \to f_af_b
\eeq
one finds the $D\bar D$ in a highly correlated $C$ odd $P$ wave configuration. Bose-Einstein statistics forbids the transformation $D^0\bar{D}^0\to D^0D^0/\bar{D}^0\bar{D}^0$ although correlated oscillations are possible. 
Therefore one wants to measure final states $f_a$ and $f_b$ that can be fed by both 
{\em coherent} sum of $D^0$ \& $\bar D^0$ or $D_+$ \& $D_-$ or $D_1$ \& $D_2$: 
\begin{eqnarray}
&&\nonumber{\rm BR}(\psi''(3770)\to D^0\bar D^0/D_+D_-/D_1D_2 \to f_af_b)\simeq
\\
&&{\rm BR}(D\to f_a){\rm BR}(D\to f_b)\times
\left[(2 - x^2_D+y^2_D)\left|\bar{\rho}(f_a)-\bar{\rho}(f_b)\right|^2+\left(x^2_D+y^2_D\right)\left|\frac{p}{q}\right|^2\left|1-\frac{q}{p}\bar{\rho}(f_a)\frac{q}{p}\bar{\rho}(f_b)\right|^2\right] \; . 
 \label{eq:fafb}
\end{eqnarray}

This rate is very small due to small values of ${\rm BR}(D\to f_a)$ and ${\rm BR}(D\to f_b)$. However there are several reasons 
to consider them:   
\begin{itemize} 
\item 
CP is violated by the {\em existence} in time integrated rates of final states that both  have CP {\em even} or 
{\em odd} quantum  numbers in $e^+e^- \to (f_a)(f_b)$, see Eq.(\ref{eq:fafb}), which provides a distinct statistical advantage over the measurement of CP asymmetries as a smaller number of events suffices. 
\item
One can{\em not} measure some phases in any other way. 
\item 
The rate of $e^+e^- \to \psi''(3770) \to D^0\bar D^0/D_+D_-/D_1D_2 \to f_af_b$ have small backgrounds.  
\item
The intermediate $\psi''(3770)$ state is a CP {\it even} state. In principle even a single event where the final state is even 
from the decay of a $D^0 \bar D^0$ pair establishes CP violation. 
 
\item 
Such CP violation can occur either due to the existence of {\it direct} or {\it indirect} CP violating dynamics  or both. 

\item 
The existence of the transition where {\it both} $f_a$ and $f_b$ are either CP even or CP odd proves the existence of CP violation in the dynamics regardless of whether $f_a$ and $f_b$ are the same or different final states.        
\item 
Indirect CP violation in neutral  kaon decays has been manifested in a much larger asymmetry in 
$K_L \to \pi^+ \pi^- e^+e^-$ decays through a final state distribution of the $\pi^+ - \pi^-$ plane vs. the 
$e^+-e^-$ plane -- at the price of the much smaller rates! Something might happen for final states 
distributions in neutral $D$ decays in $D \to h^+h^- l^+l^-$ or $D\to K^+K^- \pi^+\pi^-$ as discussed  
in the next section.

\end{itemize}
As a consequence of EPR correlations \cite{EPR}, tagging $D_+$ through final states $h^+h^-$ -- like $\pi^+\pi^-$ or $K^+K^-$ -- one gets a $D_-$ beam with some purity; likewise tagging $D_-$ through final states $K_S\phi$ one gets a $D_+$ beam. However, the correlation between the two eigenstates last only till one of them decays. After that the other eigenstate that has not decayed is prone to oscillations and the time evolution of the state has to be carefully studied. If both $f_a$ and $f_b$ are CP even [odd] it can be clearly inferred that, {\em at the time of the decay}, it was a $D_+ [D_-]$ which decayed first and the correlated state which was $D_-[D_+]$  oscillated in to the other CP eigenstate of $D^{neut}$ hence displaying the existence of indirect CP violating dynamics. Of course, we are assuming here that the effects of direct CP violation is tiny in comparison.

For the case of the $D_SD_L$ complex produced at threshold, tagging thorough final decay products is no longer trivial as in the case of the kaons since the mass eigenstates are no longer CP eigenstates to any good approximation. Experimental results allow for $0.8<|q/p|<1$ within about a 1$\sigma$ range. This, however, does not mean that an analysis on the basis on mass eigenstates is impossible. A good measure of the CP impurity in the mass eigenstate can possibly allow tagging of the mass eigenstate based on a statistical analysis.

In $e^+e^-$ annihilation close to charm threshold one can use correlations between 
$D_L$ and $D_S$ transitions like $l^+l^+X$ vs. $l^-l^-X$ and $l^+h_1h_2X$ vs. 
$l^-h_1h_2X$ with{\em out} a time filter. The fact that both $D_L$ and $D_S$ can contain sizable 
even and odd CP components leads to complex time evolutions as:
\begin{eqnarray}
\Gamma\left(\left(D^0\bar{D}^0\right)_{C=-}\to l^+_{t_1}l^+_{t_2}X\right)\propto e^{-\Gamma(t_1+t_2)}\left|\frac{q}{p}\right|\sin^2\frac{x^{}_D}{2}\bar{\Gamma}(t_1-t_2)\\
\Gamma\left(\left(D^0\bar{D}^0\right)_{C=-}\to l^-_{t_1}l^-_{t_2}X\right)\propto e^{-\Gamma(t_1+t_2)}\left|\frac{p}{q}\right|\sin^2\frac{x^{}_D}{2}\bar{\Gamma}(t_1-t_2)
\end{eqnarray}
It is also quite informative to study the time evolution of the states $[l^-X]_{t_1}[K_S\phi/h^+h^-]_{t_2}$ vs. $[l^+X]_{t_1}[K_S\phi/h^+h^-]_{t_2}$ or 
$[l^-X]_{t_1}[K^+\pi^-]_{t_2}$ vs. $[l^+X]_{t_1}[K^-\pi^+]_{t_2}$ from a pair of correlated $D^0\bar{D}^0$ mesons, as these show a different dependence on $x^{}_D$.  The asymmetry is proportional to $\sin x^{}_D(t_1-t_2)/\tau_{D^0}$ as the neutral D meson pair is produced in a C odd wave in a $e^+e^-$ machine and hence vanishes when integrated over $t_1$ and $t_2$. However there is one way out. If one studies
\begin{equation}
e^+e^-\to D^0\bar{D}^0\gamma
\end{equation}
then the correlated $D^0\bar{D}^0$ are produced in the C {\em even} state. The time dependent asymmetry is now proportional to $\sin x^{}_D(t_1+t_2)/\tau_{D^0}$ and hence does not vanish when integrated over $t_1$ and $t_2$.
The asymmetry is proportional to $$2x^{}_D\Im\left(\frac{q}{p}\bar{\rho}(f)\right)$$ 
which is the same as that for incoherently produced neutral $D$ meson.

The $e^+e^-$ machines at the $B$ factories currently run at the $\Upsilon(4S)$ resonance generating a large number of neutral $B$ mesons. As mentioned before, the decays of these $B$ mesons can be used as a ``factory'' to generate neutral $D$ mesons in their flavour eigenstate using the charge of the associate $D_s^{(*)}$ produced as a flavour tag. In such an analysis the time dependent CP asymmetry can be measured although the neutral $D$ mesons that are produced in this manner are not very highly boosted, unlike in hadron, machines and hence this allows access to values of only $t/\tau_{D^0}\sim1$.

\section{\boldmath$D_L \to h^+h^- l^+l^-$}
\label{HHLL}

 Tempted by this enhancement in $K_L$ decays we studied the analogues transitions 
$D\to h^+ h^-l^+l^-$ with $h=\pi , K$ and $l=e,\mu$. 
Indirect CP violation in $K_L$ decays can be described by BR$(K_L \to \pi^+\pi^-) \simeq 2 \times 10^{-3}$ or 
$\Gamma (K_L \to \pi^+\pi^-)/\Gamma (K_L \to \pi^+\pi^-)   \simeq 6 \times 10^{-3}$ -- i.e., a very small effect 
(and given by CKM theory  at least as the leading source). For indirect CP violation in $D^0$ transitions 
CDF has placed a 95 \% bound of $1.4 \cdot 10^{-3} \times \langle t \rangle /\tau_{D^0}$ \cite{CDFDtoHH}. At first sight that 
 number for CP violation in $D^0$ mesons looks to be in the same ballpark for the underlying dynamics, although CKM theory cannot reach it. 
\boldmath
\subsection{The Decay Width of $D_L \to h^+h^- l^+l^-$}
\unboldmath

As in the case of kaons, the decays $D_L\to h^+h^-\gamma$  can occur through two distinct channels. Firstly, it can occur through an internal bremsstrahlung associated with the CP violating decay $D_L\to h^+h^-$. It can also have a direct emission component, mostly a M1 transition with a possible E1 component coherent with the bremsstrahlung component, both being CP conserving as shown in Eq.\ref{eq:dhhg}. In $e^+e^- \to D \bar D$ -- in particular, close to charm threshold -- one finds that the final state has to carry 
even CP parity with very high accuracy due to dominance of {\em one} over {\em two} $\gamma ^*$ 
production. Therefore the pair of neutral $D$ mesons hadronize as $D_LD_S$ as for $e^+e^- \to K_S K_L$. 
First we consider  
\beq
D_L\xrightarrow{\cancel{\rm CP}}h^+h^-\xrightarrow{\rm IB}h^+h^-\gamma{\rm~and~}D_L\xrightarrow{\rm M1,E1}h^+h^-\gamma , 
\label{eq:dhhg}
\eeq
where $D_L \to h^+h^-$ can be produced from indirect and direct CP violation. 
The final state contains both a CP odd state created by the M1 transition and a CP even state due to the IB component. However, as long a the polarization of the final state photon is not observed, this interference cannot be measured \cite{SW}. The only way to observe this interference is to take the photon off-shell in which case the information about CP and T odd polarization of the virtual photon is retained by the final state leptons and the asymmetry manifests itself as an angular correlation between the $l^+l^-$  and $h^+h^-$ planes. The photon polarization can be probed best for {\em off}-shell photons 
\beq
D_L \to h^+h^-\gamma^* \to h^+h^- l^+l^- \; ; 
\eeq
in addition to the IB, E1 and M1 amplitudes one also has to include the charge radius term 
$D_L \to D_S l^+l^- \to h^+h^-l^+l^-$ which contributes to the decay rate but not to the asymmetry. In what follows we will follow the notation of Sehgal and Wanninger in \cite{SW}. The amplitude for $D_L \to h^+h^- l^+l^-$ is described by 
\begin{eqnarray}
\nonumber T(D_L &\to& h^+ h^- l^+ l^-)= e |T(D_S \to h^+ h^-)| \left[\frac{g^{}_{\rm P}}{m_D^2}\left[k^2P_\mu-(P\cdot k)k_\mu\right]\frac{1}{k^2-2P\cdot k}+\frac{g^{}_{\rm E1}}{m_D^4}\left[(P.k)p_{+\mu}-(p_+\cdot k)P_\mu\right]\right.\\
&&+\left.g^{}_{\rm BR} \left( 
\frac{p_+^{\mu}}{p_+ \cdot k} - \frac{p_-^{\mu}}{p_- \cdot k}\right) 
+ \frac{g^{}_{\rm M1}}{M_D^4} \epsilon _{\mu \nu \alpha \beta} k^{\nu} p_+^{\alpha}p_-^{\beta} 
\right] 
\frac{e}{k^2} \bar u(k_-)\gamma _{\mu}v(k_+)  \; \; 
\end{eqnarray}
where  $k= k_+ + k_-$ is the sum of the final state leptons' momenta, $p_+$ and $p_-$ the momenta of the $h^+$ and $h^-$ respectively and $P$ is the momentum of the incoming $D$ meson.
The differential decay rate is given by

\begin{eqnarray}
\nonumber \frac{\Gamma(D_L\to h^+h^-l^+l^-)}{\Gamma(D_S\to\pi^+\pi^-)}=\frac{\alpha ^2}{16\pi ^2\lambda^{\frac{1}{2}}\left(1,\mu ^2,\mu ^2\right)}\int _{4\nu ^2}^{(1-2\mu )^2}\rm{dy}\,\frac{\lambda^{\frac{1}{2}} \left[y,\nu ^2,\nu ^2\right]}{y^2}\int _{4\mu ^2}^{\left(1-\sqrt{y}\right)^2}\rm{dx}\,\lambda^{\frac{1}{2}}(1,x,y)\int_0^{2\pi }  \frac{\rm{d\Phi}}{2\pi}\,F(x,y,\Phi)\\\label{eq:BR}
\end{eqnarray}

Here $x=(p_++p_-)^2/m_D^2$ is the normalized invariant mass of the final state mesons $h=\pi/K$, $y=(k_++k^-)/m_D^2$ is the normalized invariant mass of the lepton pair $l=e/\mu$ and $\Phi$ is the angle between the plane containing the meson pair and the plane containing the lepton pair. The normalized pion and lepton mass is given by

\beq
\mu=\frac{m_h}{m_D},~~~\nu=\frac{m_l}{m_D}
\eeq
The function $F(x,y,\Phi)$ can be split into three distinct parts \cite{SW}. \\

\noindent$F_1(x,y,\Phi)$:
The terms in this part is proportional to either $\sin^2\Phi$ or $\cos^2\Phi$ and contributes to the decay rate and not the asymmetry, hence they contribute only to $\Gamma_1$ and $\Gamma_2$ in Eq.\ref{eq:gamma123}. \\

\noindent $F_2(x,y,\Phi)$: The terms in this part contribute to the asymmetry only and not the decay rate  and hence contributes to $\Gamma_3$ only in Eq.\ref{eq:gamma123}, as they are proportional to $\sin2\Phi$. \\

\noindent$F_3(x,y,\Phi)$: The part proportional to $\cos2\Phi$ which contributes to neither of the above.\\

In calculating the decay rate, terms proportional to the lepton mass have been ignored which holds good for the electrons in the final state. For muons this can introduce an error of at most $O(5\%)$. Our estimate of the branching fractions using parameter ranges that have been explained below, put them to
\begin{eqnarray}
\nonumber&&{\rm BR}(D\to \pi^+\pi^- l^+l^-)\sim 10^{-9}\\
&&{\rm BR}(D\to K^+K^- l^+l^-)\sim 10^{-10} - 10^{-9}
\end{eqnarray}

\boldmath
\subsection{The CP Asymmetry Parameter in $D_L\to h^+h^-l^+l^-$}
\unboldmath
\label{GACP}
The interference of the CP violating IB and E1 amplitude with the CP conserving M1 amplitudes will yield a {\em circularly} polarized 
photon and leads to a triple correlation between the hadrons' momenta and the photon polarization 
\beq 
P^{\gamma}_{\perp} = \langle \vec \epsilon _{\gamma} \cdot (\vec p_{h^+}\times \vec p_{h^-})\rangle \; , 
\eeq 
which is CP {\em odd}. The CP violating effect appears as a
 correlation between the $e^+e^-$ 
  and $h^+ h^-$ planes. The differential decay rate with respect to $\Phi$ can be written as:
  \beq 
  \frac{{\rm d}}{{\rm d}\Phi} \Gamma (D_L\to h^+ h^- l^+ l^-) = 
  \Gamma _1\; {\cos}^2\Phi + \Gamma _2\;{\sin}^2\Phi + 
  \Gamma _3\; {\cos}\Phi \;{\sin}\Phi. 
  \label{eq:gamma123}
  \eeq 
  It is easy to see that the term proportional to $\cos\Phi\sin\Phi$ 
  changes sign under CP and T; 
  $\Gamma _3$ thus represents CP violation. Non zero asymmetry is implied by $\int\int\int F_2(x,y,\Phi) dxdyd\Phi\ne0$ with the integral in $\Phi$ done over the difference of the unions of opposing quadrants. Thus, the asymmetry parameter is defined as

  \beq
A_{\rm T}^D=\frac{\left[\left(\int_0^\frac{\pi}{2}+\int_\pi^\frac{3\pi}{2}\right)-\left(\int^\pi_\frac{\pi}{2}+\int^{2\pi}_\frac{3\pi}{2}\right)\right]\frac{d\Gamma}{d\Phi}d\Phi}{\int_0^{2\pi} \frac{d\Gamma}{d\Phi}d\Phi} = 
  \frac{2\Gamma _3}{\pi (\Gamma _1 + \Gamma _2)} .
  \label{ASEHGAL}  
  \eeq

A detailed look at the form of $F(x,y,\Phi)$ in \cite{SW} will make it clear that the asymmetry comes from the interference of the CP conserving M1 amplitude and CP violating E1 and bremsstrahlung components. One should note that the CP/T odd correlation is controlled by $\eta ^D_{h^+ h^-}$, which enters through $g^{}_{\rm BR}$, as we shall see later and is given by
\beq
\eta ^D_{h^+ h^-} = \frac{T(D_L\to h^+h^-)}{T(D_S\to h^+h^-)}
\eeq
which can arise from both 
indirect and direct CP violation; also one should understand the dynamical situation is very different from 
$K_L$ decays: 
\begin{itemize}
\item 
Unlike the known situation in $K$ decays direct and indirect CP violation in $D$ transitions are 
of similar strength within the SM and are weak. In ND they could also be of similar strength and sizable -- 
they could reach the upper experimental bounds listed in Eq.(\ref{DOSCDATA}). In one 
class of ND, namely LHT, one finds that indirect CP violation could be large, yet direct CP asymmetries weak. On the other hand, models with multiple Higgs doublets can contribute sizably to both direct and indirect CP asymmetries.
\item
One gets more channels where CP/T odd correlations can exist and even measured, namely 
$D_L \to \pi^+\pi^- l^+l^-$, $K^+K^- l^+l^-$ with $l=e$, $\mu$. 
One can compare $\eta ^D_{\pi^+ \pi^-}$ vs. $\eta ^D_{K^+ K^-}$: a  
difference would show direct CP violation, but $\eta ^D_{\pi^+ \pi^-} = \eta ^D_{K^+ K^-}$ does not prove 
the absence of direct CP violation. However, one can compare $\eta ^D_{\pi^+ \pi^-}$ vs. $\eta ^D_{\pi^0 \pi^0}$ 
as usual. One could also study $e^+e^- \to D^0 \bar D^0 \to D_S +[K_Sf_0(980)]$ and probe 
$D_S \to K_S\pi^0 l^+l^-$, $K_S\rho^0 l^+l^-$ for a CP/T odd correlation. 
\item
While the correlation is controlled by $\eta ^D_{h^+ h^-}$, it is also governed by other factors that depend on the impact of strong forces. For $K_L \to \pi \pi \gamma^{(*)}$ 
and $K_L \to K_S e^+e^-$ one can apply chiral dynamics with small  spaces, but there are more problems 
and uncertainties for $D$ transitions. 
\end{itemize}

We know that the differences between $D^0 \to h^+h^-$ and $\bar D^0 \to h^+h^-$ are small. Indirect 
CP asymmetry are suppressed by $x^{}_D$ and $y^{}_D$ of about 0.005 - 0.01. In preparing $D_L$ samples, one 
can find indirect CP violation in $D_L \to h^+h^-$ without this penalty. One might enhance the strength of  CP violation in $D_L \to h^+h^- l^+ l^-$ from ND at the price of much smaller branching ratios -- if the strong 
forces collaborating as they did for $K_L \to \pi^+\pi^- e^+e^-$.

\boldmath
\subsection{A Note on the Short Distance contribution to $D_L\to h^+h^-l^+l^-$}
\unboldmath

In addition to the contributions discussed in the previous sections, there can be a short distance contribution to $D_L\to h^+h^-l^+l^-$ analogous to similar effects in $K_L\to \pi^+\pi^-e^+e^-$ as discussed in \cite{HS}. The short distance contribution can be classified into three types. There can be vector and an axial current components stemming from a local interactions of the current $\bar{u}\gamma_\mu(1-\gamma_5)c$ with $\bar{l}\gamma_\mu l$ and $\bar{l}\gamma_\mu \gamma_5l$ respectively. These amplitudes can possibly contribute to both the decay width and the CP asymmetry. Within the standard model this contributions will be highly CKM suppressed and a rough estimate clearly shows that it will be several orders of magnitude smaller than any of the contributions discussed above. The third component leads to the CP violating E1 amplitude that has been discussed above. This component is also CKM suppressed within the SM.  

ND like LHT can have large effects on these amplitudes and can also lead to enhancements of several orders of magnitudes over the SM in certain parts of the parameter space. This is not unusual as such has been seen even for other short distance operators studied previously for other decay channels of the neutral charm meson. However, even with such enhancements, long distance effects remain the dominant contribution to this decay channel. 

\subsection{Estimation of Parameters}
\label{PARAM}

To complete the analysis we have to estimate the magnitude and phases of the following five parameters from experimental measurements.

{\bf (i) The decay constant \boldmath $f_S^D$:}
The decay constant $f_S^D$ is defined by 
\beq
\Gamma(D_S\to h^+h^-)=\frac{|f_S^{D(h)}|^2}{16\pi m_\m}\left(1-\frac{4m_h^2}{m_D^2}\right)
\eeq
we will only require the phase information of $f_S^D$.\\

{\bf (ii) The bremsstrahlung parameter \boldmath $g^{}_{\rm BR}$:}
$g^{}_{\rm BR}$ is defined as
\beq
g^{}_{\rm BR}=\eta_{h^+h^-}^D\frac{f_S^D}{|f_S^D|}, ~~ \arg\left(g^{}_{\rm BR}\right)=\Phi_\pm^{D}+\delta_0\label{eq:gbr}
\eeq
where
\beq
\eta_{h^+h^-}^{D(h)}\equiv\frac{\langle h^+h^-|H_W|D_L\rangle}{\langle h^+h^-|H_W|D_S\rangle}=\epsilon^{}_D+\epsilon_D^\prime, ~~ \arg\left({\eta_{h^+h^-}^{D(h)}}\right)\equiv\Phi_\pm^{D(h)}\label{eq:etapm}
\eeq
Assuming that the magnitude of $\eta^D_{h^+h^-}$ is driven purely by oscillations and the effect of direct CP violation is tiny we get
\beq
\eta_{h^+h^-}^D\simeq\epsilon^{}_D\simeq\frac{1}{2}\left(1-\frac{q}{p}\right)
\eeq
which is an approximation good to about $O(10\%)$ since in the adopted CKM phase convention the assumption that $T(D^0\to f) \simeq\bar{T}(\bar{D}^0\to f)$ holds good \cite{DKdual}. Taking the 95\% CL values from HFAG\cite{HFAGCHARM},
\beq
\left|\frac{q}{p}\right|\in\{0.6,1.28\}, ~ \phi\in\{-27.5^o,8.4^o\} \implies \left|\eta_\pm^D\right|\in\{0,0.3\},~\Phi_\pm\in\{-\pi,\pi\}
\eeq
The phase $\delta_0$ in Eq.\ref{eq:gbr} is the $hh$ scattering phase in the $I=J=0$ channel at $\sqrt{s}=m_D$. \\

{\bf (iii) The M1 parameter \boldmath $g^{}_{\rm M1}$:}
A measurement of the direct emission component of $D_L\to h^+h^-\gamma$ is necessary to estimate $g^{}_{\rm M1}$
\beq
\frac{ \Gamma(D_L\to h^+h^-\gamma;\omega>\omega_{min})}{\Gamma(D_S\to h^+h^-)}=\frac{\alpha}{\pi}(g^{}_{\rm M1})^2\int^{\omega_{max}}_{\omega_{min}}\frac{\omega^3d\omega}{m_D^4}\frac{\beta^3}{6\beta_0}\left(1-\frac{2\omega}{m_D}\right)=(g^{}_{\rm M1})^2I_{\rm M1}(m_h)
\eeq
where
\beq
\beta=\left(1-\frac{4m_h^2}{m_D^2-2m_D\omega}\right)^\frac{1}{2},~~\beta_0=\beta\arrowvert_{\omega=0},~~\omega_{max}=m_D\frac{\beta_0^2}{2}
\eeq
With a cutoff of $\omega_{min}=20 MeV$, $I_{\rm M1}(m_h)$ is of $O(10^{-6})$ for $h=\pi$ and $O(10^{-7})$ for $h=K$. Hence, although we cannot calculate $g^{}_{\rm M1}$ at this point for the lack of experimental data, we can estimate it to be of $O(1)$ -- $O(10^{-1})$. The phase of $g^{}_{\rm M1}$ is $\delta_1(s_h)+ \pi/2$, where $\delta_1(s_h)$ the $hh$ scattering phase in the vector channel at a c.m. energy of $\sqrt{s_\pi}$ and the phase shift of $\pi/2$ has to be introduced to allow for the phase convention used for $g^{}_{\rm BR}$ \cite{SW}.\\

{\bf (iv) The E1 parameter \boldmath $g^{}_{\rm E1}$:}
The E1 component is distinct from the IB component and is a CP even component in the $D_L\to h^+h^-\gamma$ amplitude. The phase of $g^{}_{\rm E1}$, however, depends on whether it comes from the CP impurity in the $D_L$ wavefuntion or from an intrinsic CP violating dynamics in the CP odd part of $D_L$. In the former case the phase of $g^{}_{\rm E1}$ relative to that of $g^{}_{\rm M1}$ is given by 
\beq
\arg\left(\frac{g^{}_{\rm E1}}{g^{}_{\rm M1}}\right)=\Phi_\epsilon\approx\Phi_{\pm}^D-\frac{\pi}{2}
\eeq
In the latter case $g^{}_{\rm E1}$ is out of phase with $g^{}_{\rm M1}$ and hence does not contribute to the asymmetry. As in \cite{SW} we will continue to use the first scenario and not the second. Also, as in the case of the kaons, it is logical to assume that $g^{}_{\rm E1}$ will be tiny compared $g^{}_{\rm M1}$ and in the lack of any evidence to the contrary we shall maintain the same. It is unlikely that ND contributions to the E1 amplitude can enhance $g_{E1}$ significantly. Hence we shall take
\beq
\left|\frac{g^{}_{\rm E1}}{g^{}_{\rm M1}}\right|=0.05
\eeq

{\bf (v) The charge radius contribution parameter \boldmath $g^{}_{\rm P}$:}
The amplitude connected to the charge radius does not contribute to the asymmetry but contributes to the total decay rate. This parameter depends on the estimate of the $D$ meson charge radius, which is estimated to be twice as large as the neutral Kaon charge radius. Along with the fact that $m_D\sim 3.6 \times m_K$, we estimate $g^{}_{\rm P}$ to be

\beq
g^{}_{\rm P}=-\frac{1}{3}\langle R^2\rangle m_D^2\approx 7.8
\eeq
The phase of $g^{}_{\rm P}$ is $\delta_0(s_h)$ which is the $hh$ scattering phase in the $I=0$ channel at c.m energies of $\sqrt{s_h}$. It should be noted that we do not consider any modifications to $g^{}_{\rm P}$ due to the off-shell behavior of the $D_S\to h^+h^-$ amplitude as pointed out in \cite{HS}

\subsection{Numerical Results}
\label{RES}

\begin{figure}[h!]
\includegraphics[width=17cm]{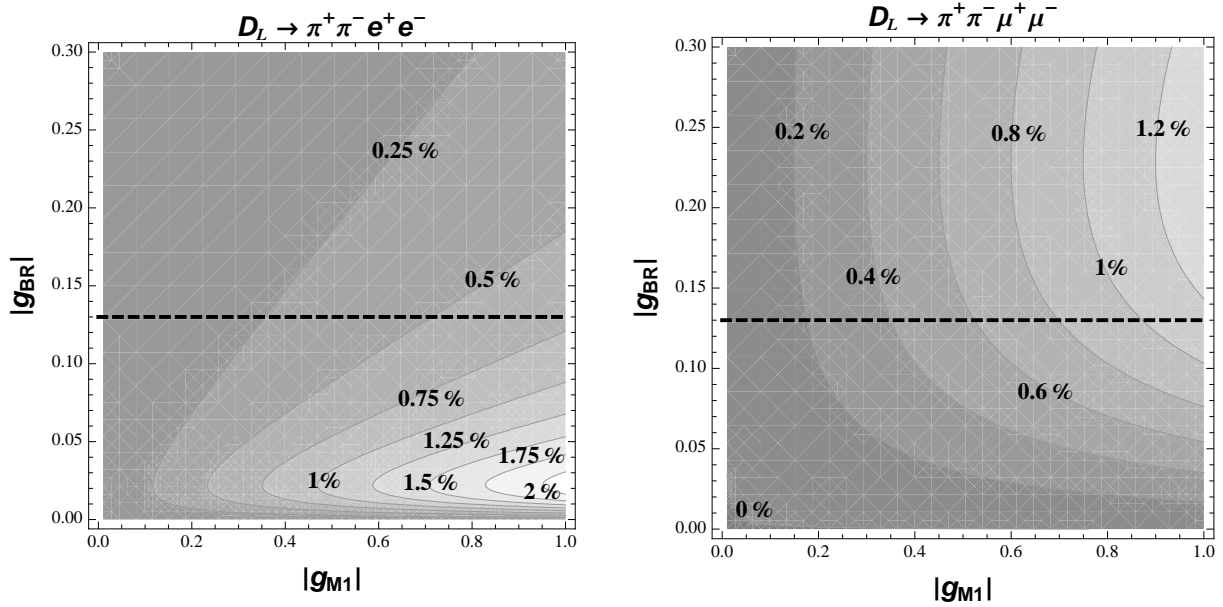}
\caption{Contour plot of the CP asymmetry parameter $A^D_{\rm T}$ with pions in the final state in the space of the couplings parametrizing the M1 and bremsstrahlung contributions. Results for both electrons and muons in the final state are shown above. The dashed line marks the upper limit on $g_{\rm BR}$ when assuming there is no direct CP violation.}
\label{fig:Dpipi}
\end{figure}
\begin{figure}[h!]
\includegraphics[width=17cm]{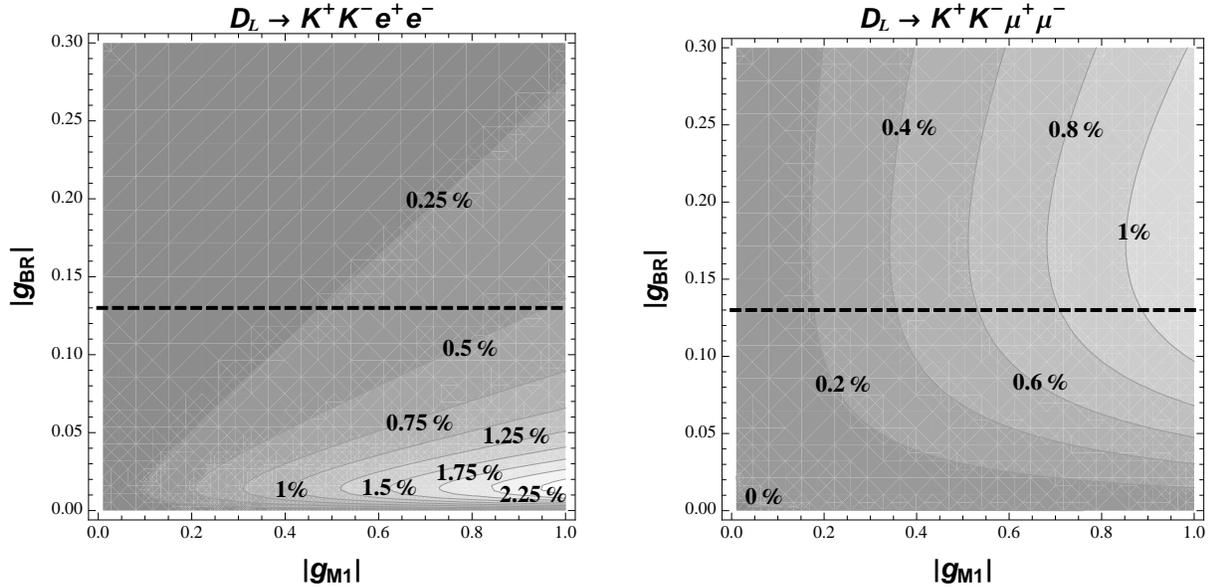}
\caption{Contour plot of the CP asymmetry parameter $A^D_{\rm T}$ with kaons in the final state in the space of the couplings parametrizing the M1 and bremsstrahlung contributions. Results for both electrons and muons in the final state are shown above. The dashed line marks the upper limit on $g_{\rm BR}$ when assuming there is no direct CP violation.}
\label{fig:DKK}
\end{figure}

From the previous section it should be quite evident that a significant degree of uncertainty plagues the estimation of the required parameters. An exact (or near exact) calculation of this asymmetry requires far more experimental input than we currently have. The largest uncertainty, by far, comes from the estimation of the charge radius parameter $g^{}_{\rm P}$ which is only a theoretical estimate of a purely long distance contribution and hence way beyond our theoretical control. There are also large uncertainties in the estimation of both the magnitude and the phase of $g^{}_{\rm BR}$, the constant parametrizing the bremsstrahlung contribution to the amplitude. The uncertainty stems from the uncertainty in the measurement of $|q/p|$ and $\phi$. The constant $g^{}_{\rm M1}$ can only be determined empirically. As for $g^{}_{\rm E1}$, the contribution of the E1 channel is small for both the decay rate and the asymmetry, and hence the uncertainty stemming from it is not significant. 

Despite these large uncertainties, it is possible to set an upper bound on the size of the asymmetry one can expect in these decay channels. In the analysis to follow we shall show how the upper limit on the asymmetry varies with the uncertainty in the magnitude of each parameter. As for the phases, since there is no concrete way of constraining their range, to get the upper limit on the asymmetry, we shall assume maximal phase contribution to the asymmetry which in mathematical terms means, given
\bea
A_{\rm T}^D&=&A_1\cos\Theta_1+A_2\cos\Theta_2\\
\Theta_1&=&\arg(g^{}_{\rm M1}g^{*}_{\rm BR})=(\Phi_\pm+\delta_0-\delta_1-\frac{\pi}{2})\mod\pi\\
\Theta_2&=&\arg(g^{}_{\rm M1}g^{*}_{\rm E1})=(\Phi_\pm-\frac{\pi}{2})\mod\pi
\eea
we shall assume $\Theta_1\approx 0/\pi$ and $\Theta_2\approx 0/\pi$. It should also be noted that we are not varying the parameter $g^{}_{\rm P}$ due to our lack of theoretical control over it. Suffice it to say that a smaller than estimated magnitude of $g^{}_{\rm P}$ will raise the maximum value of the asymmetry from $O(1\%)$ to $\sim 10\%$, the phase contribution remaining maximal.

In Fig.\ref{fig:Dpipi} we plot the variation of the asymmetry over the range of possible values of $g^{}_{\rm M1}$ and $g^{}_{\rm BR}$. For electron pairs in the final state the asymmetry can be as large as 2\%. However, it should be noted relatively large asymmetry is found for relatively smaller values of $g^{}_{\rm BR}$ and large values of $g^{}_{\rm M1}$. This is contrary to the plot in Fig.\ref{fig:Dpipi} that refers to muon pairs in the final state where the larger asymmetry occurs in the region of larger $g^{}_{\rm BR}$ and $g^{}_{\rm M1}$. Also, the maximum asymmetry in the decay channel with muons in the final state is smaller by orders of magnitude as compared to the ones with electrons in the final state. These, in general, seem to hold good for kaon pairs in the final state as seen in Fig.\ref{fig:DKK}.

Both the fact that the asymmetry is much lower in the case of muonic final states and its dependence on $g^{}_{\rm M1}$ and $g^{}_{\rm BR}$ can be understood from the finiteness of the muon mass as compared to the electron mass. The bremsstrahlung component of the amplitude is sensitive to the mass of the final state leptons and there is a natural cutoff for this contribution from the muonic mass which serves to naturally curtail the infrared sensitivity of this amplitude at low leptonic invariant mass. However this is absent for the electrons and large infrared contributions are seen in the bremsstrahlung component and its interference with other amplitudes. 

The numerator in the asymmetry depends linearly on the bremsstrahlung contribution, but the denominator depends quadratically on it. Hence for the electrons the asymmetry peaks and lower values of $g^{}_{\rm BR}$ before the quadratic dependence in the denominator takes over and reduces the asymmetry. On the other hand, for muonic final states this does not happen as the bremsstrahlung contribution faces a natural infrared cutoff at the dimuon mass.

From Fig.\ref{fig:Dpipi} and Fig.\ref{fig:DKK} it is evident that assuming that there is no source of direct CP violation in charm dynamics does not affect the possible size of the asymmetry to a large extent. While for the case of electrons in the final state, the region of maximum asymmetry lies well within the upper bound set by the dashed line, for muonic final states regions of larger asymmetries are not completely excluded.

\boldmath
\subsection{$e^+e^- \to [h^+h^-][l^+l^-h^+h^-]$, $[K_S\phi][l^+l^-h^+h^-]$}
\unboldmath

While even the existence of $e^+e^- \to [D^0\bar{D}^0]_{C=-}\to [h^+h^-][l^+l^-h^+h^-]$ is a signal for CP violation the same is not true for $e^+e^- \to [D^0\bar{D}^0]_{C=-}\to[K_S\phi][l^+l^-h^+h^-]$. However, a detailed study of correlations in both the final states is necessary. Since the final states are common to $D^0$ and $\bar{D}^0$, the integrated partial width of the correlated states is given by Eq.\ref{eq:fafb}

\section{Other four body decays with a lepton pair in the final state}
\label{LLFINAL}
Within the SM and the KM ansatz, CP violation can manifest itself in Cabibbo suppressed decays of charmed mesons. However, as mentioned earlier, there are some subtle exceptions. CP violation can exist in the DCS decay $D^0\to K^+\pi^-$ vs. $\bar{D}^0\to K^-\pi^+$ and the apparently Cabibbo allowed decays $D^\pm \to K_S\pi^\pm$.

\begin{description}

\item {\boldmath $D^0 \to K^+\pi^-l^+l^-$ vs. $\bar D^0 \to K^-\pi^+l^+l^-$}\\
CP violation can manifest itself in $D^0\to K^+\pi^-$ vs. $\bar{D}^0\to K^-\pi^+$ through the interference of $\Delta C=1$ dynamics with time dependent CP asymmetry as mention earlier. An enhanced asymmetry can possibly be observed as in the case of $D^{neut}\to h^+h^-l^+l^-$, as discussed above, through the measurement of a T odd correlation although the same price has to be paid: a much reduced branching fraction.

\item {\boldmath $D^+ \to K_S h^+l^+l^-$ vs. $D^- \to K_S h^-l^+l^-$}\\
For $D^\pm \to K_S\pi^\pm$, an apparently Cabibbo allowed mode, the final state can be accessed through a DCS reaction through $K^0 - \bar{K}^0$ oscillations and the two amplitudes interfere, opening up possibilities of CP violation even within the SM and, more importantly, allows for possible ND contribution as mentioned earlier. This can also lead to a T odd correlation in the comparison of $D^\pm \to K_S \pi^\pm l^+l^-$.

In the case of $D^\pm \to K_S K^\pm$, a Cabibbo suppressed channel CP violation is manifested only through the final state and is quantified by $\Re(\epsilon_K)$. This effect will be reflected in the comparison of $D^\pm \to K_S K^\pm l^+l^-$. ND cannot leave any impact here.

\end{description}

\boldmath
\section{\boldmath $D^{\rm neut} \to h l\nu$ and $D\to h \nu \bar \nu$}
\unboldmath
\label{HLNU}

\boldmath
\subsection{$D^0 \to l^-\bar \nu h^+$ vs. $\bar D^0 \to l^+ \nu h^-$}
\unboldmath

The rate of both $D^0 \to l^-\bar \nu h^+$ and $\bar D^0 \to l^+ \nu h^-$ are given by the {\em same} factor 
$e^{-\bar{\Gamma} t}$, but suppressed by the tiny factor $(x^2_D + y^2_D)/2$. Their asymmetry is 
given by $\frac{q^4 - p^4}{q^4 + p^4}$ independent of their time of decays. Yet the experimental 
bound on $|q/p|$ allows the range from about 0.73 to 1.1 and allows the $A_{\rm sl}^{\rm CP}$ to range from 
-0.56 to 0.19 -- i.e., a possible large asymmetry. Furthermore {\em non-ad-hoc} models of ND -- like LHT -- allow 
a significant part of this range, in particular if it produces a sizable, but not leading part of the observed value of $x^{}_D$. The same is true for models with non-minimal Higgs sectors.

As stated before this asymmetry does not depend on the time of decay of the neutral charm meson. Yet two comments should be mentioned here: 
\begin{itemize}
\item 
The data, on which the observation of such an asymmetry are based, should show the `known' time dependance 
for both $D^0$ and $\bar D^0$ transitions. 
\item 
If ND produces `wrong sign' decay through $\Delta C=1$ dynamics, it would generate different time dependance of 
$D^0 \to l^- X$ vs. $\bar D^0 \to l^+X$. 

\end{itemize}
It is crucial to determine whether the neutral meson was a $D^0$ or $\bar D^0$. The usual flavour tagging is 
done by $D^{*+} \to D^0 \pi^+$ vs. $D^{*-} \to \bar D^0 \pi^-$. Another way would be to employ 
$B^- \to D^0 D_s^{(*)-}$ and $B^+ \to \bar D^0 D_s^{(*)+}$ to tag $D^0$ vs. $\bar D^0$ as discussed before.

\boldmath
\subsection{Asymmetries in $D_{L,S} \to l^{\pm} \nu h^{\mp}$}
\unboldmath

One can consider a search for a CP asymmetry in semi-leptonic decays of $D$ mesons in analogue with kaon 
decays, where one has measured the predicted difference in $K_L \to l^{\pm}\nu \pi^{\mp}$. However the analogy 
is at best {\em only qualitative}, not quantitative. For neutral kaons one has large oscillations due to 
$\Delta M_K/\bar \Gamma _K$ and a huge difference in life times, while $||q_K/p_K| - 1| < 0.005$; for 
neutral $D$ mesons we have small values of $x^{}_D$, $y^{}_D$ in the range 0.005 - 0.01, while 
$|q_D/p_D|$ could differ from unity by 10\% or more. In $e^+e^-$ annihilation close the charm threshold one can study $e^+e^- \to D_SD_L$ denoting the two 
mass eigenstates with short and long lifetimes and make use of quantum correlations -- as one had 
for $e^+e^- \to \phi (1020) \to K_SK_L$.  Of course one has to consider the hurdle of small lifetime differences in the neutral $D$ system but can reap benefits from the possible large CP impurity in the mass eigenstates.

\boldmath
\subsection{$D \to h \nu \bar \nu$}
\unboldmath
In a previous work \cite{PBR2} we have discussed rare charm decays like $D \to l^+l^-X$ and found that LD dynamics have a 
strong tendency to overwhelm signals from ND, unless one finds CP asymmetries there. Yet the situation 
is different for $D \to X\nu \bar \nu $, since LD dynamics cannot contribute there in an appreciable way. SM dynamics can generate $D \to X\nu \bar \nu$, yet with branching ratios well beyond possible experimental reach. A simplified calculation using the Inami-Lim functions \cite{InamiLim} leads to the numbers \cite{Burd}:
\begin{equation}
{\rm BR_{SD}}(D^+\to X_u\nu\bar{\nu})\simeq 1.2\times 10^{-15},\;\;\;{\rm BR_{SD}}(D^0\to X_u\nu\bar{\nu})\simeq 5\times 10^{-16}
\end{equation}

 \begin{figure}[h!]
\subfigure[$m_{HL}=400GeV$]{\includegraphics[width=16cm]{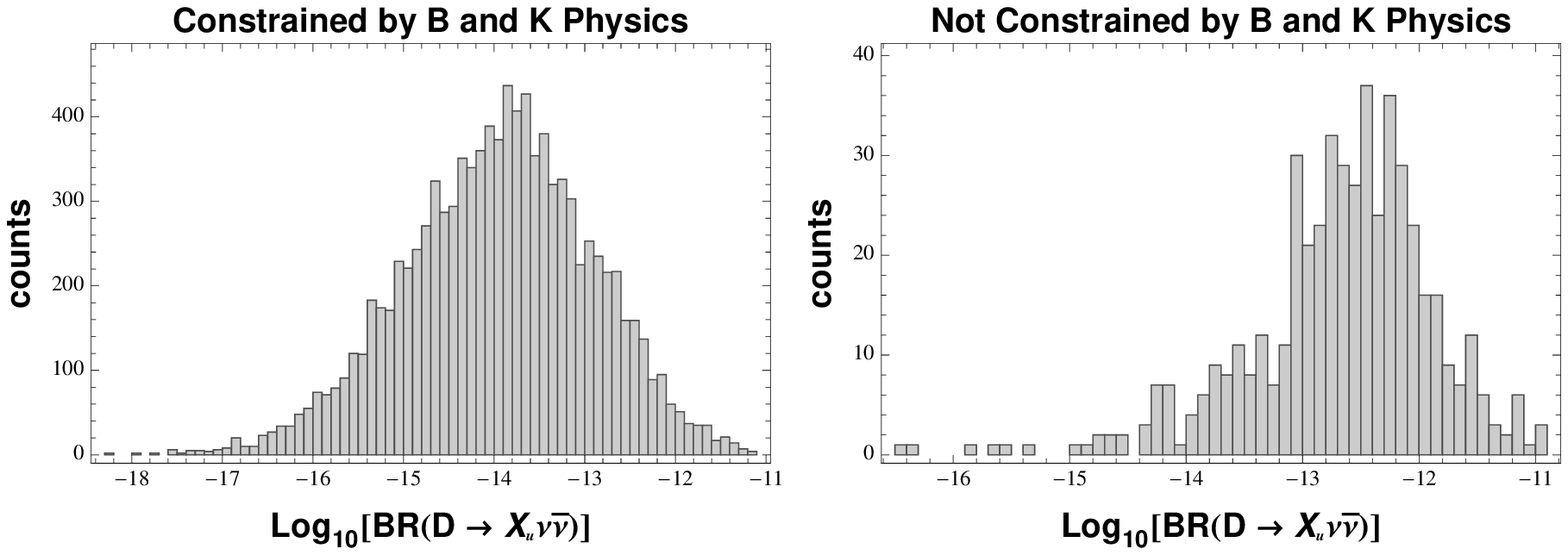}
\label{fig:Dnunu400}}
\subfigure[$m_{HL}=1100GeV$]{
\includegraphics[width=16cm]{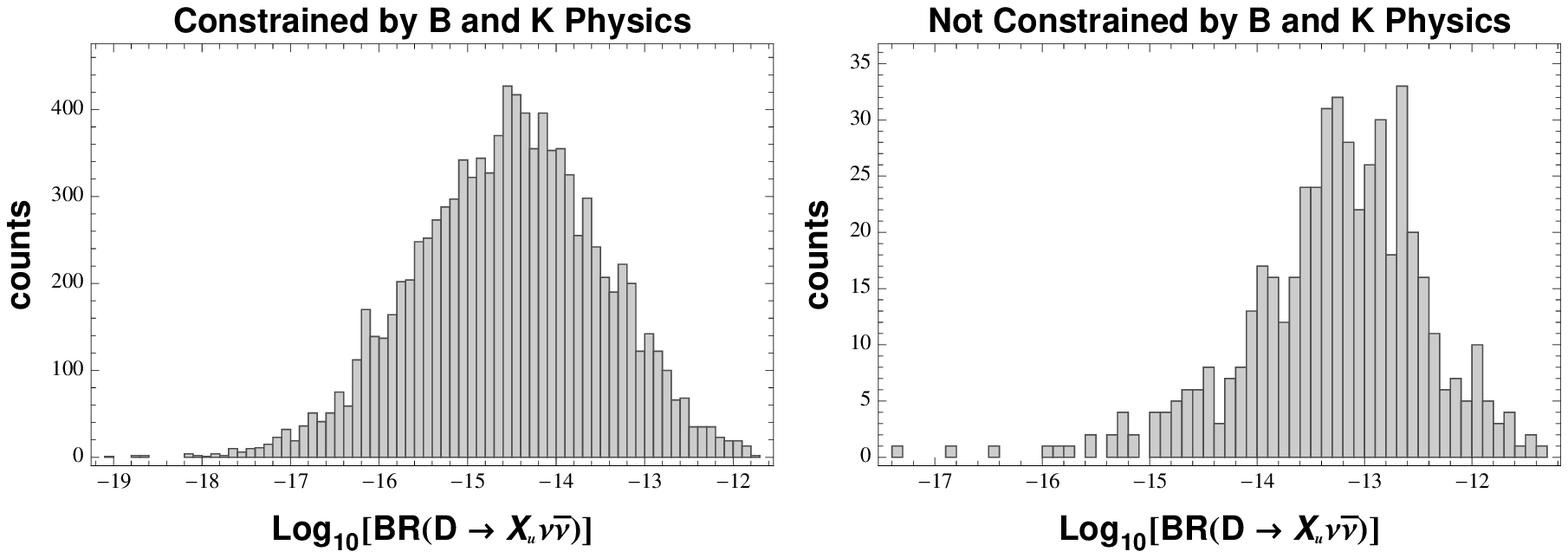}
\label{fig:Dnunu1100}
}
\caption[]{Histograms of  ${\rm BR}(D\to X_u \nu\bar{\nu})$ including LHT contributions for mirror lepton mass of  \subref{fig:Dnunu400} $m_{HL}=400GeV$ and \subref{fig:Dnunu1100} $m_{HL}=1100GeV$}
\label{fig:DnunuLHT}
\end{figure}

However, to treat this decay channel properly one has to take into account the fact that the quarks running in the loops of the penguins and the boxes are essentially light quarks and cannot be treated like the top quark loop as is done in the analogous $B$ or $K$ decay channel. To reduce theoretical uncertainties one would have to do a complete two-loop renormalization group analysis \cite{BuBu} as is done for the light quark loops in the analogous $K$  decay channel which, considering the inaccessibility of these channel to experiments, is not a quite fruitful exercise at present.

On the other hand, since these decay modes are highly suppressed in the SM and get negligible contribution from LD dynamics within the SM, they are ripe for ND intervention. Our analysis of these channels within a LHT-like\footnote{For the details of the ND c.f. sect.\ref{LHMT}} ND reveals that enhancements of $O(10^3) - O(10^4)$ is possible in a large part of the considered parameter space\footnote{A description of the parameter space can be found in \cite{PBR2,PBR1}} as can be seen in Fig.\ref{fig:DnunuLHT}. Even with such enhancements, these decay modes will remain beyond realistic experimental reach for quite some time.

Within the SM, there can be no CP violation in $D \to X\nu \bar \nu $. LHT-like models can do very little to change this as direct CP violation remains extremely tiny and mostly nonexistent in such ND.

Furthermore the mere existence of $D_L \to \pi^0 \nu \bar \nu$ would already show CP violation. At the leading order this process can occur only through $Z^0$ penguins and $W^\pm$ box contributions, both of which would be CP violating channels. The argument is clearer when one looks at the effective hamiltonian
\begin{equation}
\mathcal{H}_{eff}=c_{CPV}^{}\phi_{\pi^0}\partial_\mu\phi^{}_{D_L}\bar{\nu}\gamma^\mu\nu
\end{equation}
Under a CP transformation $\phi_{\pi^0}\to - \phi_{\pi^0}$, $\partial_\mu\phi^{}_{D_L}\to - \partial_\mu\phi^{}_{D_L}$ and $\bar{\nu}\gamma^\mu\nu\to - \bar{\nu}\gamma^\mu\nu$ making $\mathcal{H}_{eff}$ CP odd. Since there are no higher order contributions to this channel (unlike $D_L\to \pi^0e^+e^-$), in the absence of CP violation this channel is forbidden. Hence, any finite measurement of this decay mode is a clear signal of CP violation.

\section{Comments on \boldmath $D\to h_1h_2 h_3h_4$}
\label{HHHH}

Obviously there are theoretical and experimental challenges in treating four-body final states. However 
the analyses of four-body decays offer advantages too, namely there are several observables of 
CP asymmetries that can (a) offer checks on measurements and (b) can give complementary information 
about the underlining dynamics. We give some examples in non-leptonic $D$ decays. 

Let us consider $D^0\to K^+K^-\pi^+\pi^-$. As usual, a time dependent CP asymmetry can be studied here by comparing $D^0\to K^+K^-\pi^+\pi^-$ vs. $\bar{D}^0\to K^+K^-\pi^+\pi^-$. However there are other probes that can be employed very effectively here. The T odd correlation is defined as:
\beq
C_T\equiv \langle(\vec{p}_{K^+}\times(\vec{p}_{\pi^+})\times \vec{p}_{\pi^-})\rangle
\eeq
which under time reversal transforms as: $C_T\to -C_T$. However, $C_T\ne 0$ does not conclusively establish T violation as it can be driven by FSI through T invariant dynamics, time reversal being an anti-unitary operator. The effects of this FSI intervention can be resolved by measuring the conjugate T odd correlation:
\beq
\bar{C}_T\equiv \langle(\vec{p}_{K^-}\times(\vec{p}_{\pi^-})\times \vec{p}_{\pi^+})\rangle
\eeq
as $C_T\ne- \bar{C}_T$ unambiguously proves the existence of CP violation.

The effects of ND can appear in this asymmetry without having left a large impact in the partial decay rates themselves. Arguments placed in sect.\ref{D0hhll} also govern the size of the asymmetry here. A comprehensive study of this asymmetry with LHT-like ND is not presented in this paper and is left as a future task. Some interesting comments on such four-body hadronic final states can also be found in \cite{GR}.

\section{A non-ad-hoc scenario of New Dynamics -- the Little Higgs Models with T Parity }
\label{LHMT}

Flavour physics has often served as testing grounds for any ND and has played a majors role in the christening of different models as plausible candidates for dedicated searches in some confines of their allowed parameter spaces. The Little Higgs Models \cite{LH,SimHiggs,Perel,Han} are no different. These models were originally formulated to address the hierarchy problem through extensions of the SM global and gauge groups and the judicious arrangement of their couplings along with collective symmetry breaking, something which is complimentary to how Supersymmetric (SUSY) models address the same problem, through transformations of spin statistics. However, like in SUSY, a $\mathcal{Z}_2$ symmetry, called T parity \cite{CL,Low}, had to be postulated to limit contributions that could break the SM custodial $SU(2)$ symmetry which, in turn, would allow the scale of ND to be brought down to around 1 $TeV$. 

One of these class of models found favour amongst theorists for its minimal structure: the Littlest Higgs Model \cite{L2H} with T parity, having a relatively small global and gauge structure and only a few free parameters. One implementation of T parity in this model necessitates the introduction of a set of ``mirror'' fermions\cite{LHTRev,VHDVHUCKM,angph}. Although the model was built to address the hierarchy problem, to make its implementation consistent required the model to be naturally non-minimal flavour violating, a feature which is much sought after by flavour physicists. This feature then opens up avenues for possibly interesting ND features in rare decays and CP asymmetries in K, B \cite{LHTBK} and D \cite{DtoHH,PBR2,DKdual,PBR1} decays, a large part of which has been extensively studied and exists in the literature. 

What makes the flavour structure of this model particularly well studied is that a significant space of the same favourable parameter sets have been studied in $K$, $B$ and $D$ systems generating significant correlations between such ND contributions to oscillations, rare decays and CP violating parameters in all three mesonic systems. These studies not only bring together the ``up'' and the ``down'' type in the SM, but also the ``Up'' and the ``Down'' type in the mirror fermion sector.

For the $D$ meson system, it has been shown that LHT can account for all the strength of the oscillations in the neutral system \cite{DKdual}. It has also been shown that LHT can generate very little {\it direct} CP violation in the inclusive leptonic channel and two body final state hadronic channel \cite{DtoHH,PBR2,PBR1} with SM long distance contributions overshadowing contributions from LHT to decay rates in the rare decay modes. However, the story is very different for {\it indirect} CP violation where LHT can leave its mark through its contribution to the oscillation and mixing parameters. The philosophy which governs our view of ND intervention in the two, three and four body decay modes that we discuss in this work is based on the fact that in the light of the uncertainty that surrounds the SM contribution to indirect CP violation in the $D$ meson system, LHT-like\cite{DtoHH,PBR2,PBR1} ND can step in and make some or all of the contribution possible. 

What has been important in all our previous analysis on such ND in $D$ systems, and is also pertinent to the current analysis, is the fact that we do not see LHT in itself as being the solution that nature chose, but rather, when stripped of the detailed constructs of the model there lies in it a very interesting flavour structure that might be generalized by certain principles elucidated in \cite{DtoHH,PBR2,PBR1}. Such flavour structure can be generalized to a few other models that have already been postulated or can appear in other flavour structures that might be formulated in the future. As our analysis is based on general constraints and extensions of SM like flavour structures, we feel that our conclusions go beyond LHT itself and encompasses any model that might share with it similar flavour dynamics.

\section{Comments on ND with non-minimal Higgs Sectors}
\label{NONMINHIGGS} 
Having considered a class of models that can leave large impact on indirect CP violation while leaving almost no footprint on direct CP violation, it is natural to consider another class of models that can affect both. One such class of models can be built by introducing non-minimal Higgs sectors in the ND. Such models typically have extra scalar and pseudoscalar degrees of freedom and CP violation can be introduced through complex Yukawa couplings and/or mixing between the scalars and pseudoscalar higgs. 

Depending on the details of the Higgs sector, i.e., the number of Higgs doublets and the nature of the Yukawa couplings, different implementations of a non-minimal Higgs sector have different contributions to both FCNCs and CP violations in the quark sector. Varying prescriptions can be used to reduce the contribution of these new Higgs sectors to FCNCs to make them compatible with electroweak precision measurements and flavour physics constraints. Most common prescriptions lead to a MFV structure. However, it is a common misconception that in such MFV constructs there can be no new sources of CP violation\footnote{For an intriguing counter example see \cite{PichA2HDM}}. While dangerous contributions to FCNCs can be avoided by the prohibition of new sources of tree level contributions to FCNCs, CP violation beyond the CKM construction can arise from new flavour-blind phases. It is also possible to arrange for some simple symmetries in some models with multiple Higgs doublets such that FCNC contributions to the down type quarks is limited to within the experimental bounds while allowing significant contributions to the up type quarks \cite{PS}.

Unlike, the LHT-like models considered above, models with non-minimal Higgs sectors can contribute to both direct and indirect CP violations. While the presence of a charged Higgs along with possible complex phases in the Yukawa coupling can induce tree level direct CP violation, the running of these scalar/pseudoscalar degrees of freedoms in loops can induce indirect CP violation. For the case of the neutral $D$ meson such sources of indirect CP violation can saturate the experimental bounds on the oscillation parameters. Moreover, charged Higgs currents can leave tree level contribution to direct CP violation in $D^0\to\pi^+\pi^-$ and more so in the DCS decay $D^0\to K^+\pi^-$ where SM contribution is nonexistent. As we have seen above, CP violation in $D^L\to h^+h^-l^+l^-$ is driven by the CP violation is $D^0\to h^+h^-$ and hence enhancement in the latter can easily manifest itself as an enhancement in the former. A detailed study of such effects in charm physics is left as a future task.

\section{Summary}
\label{SUM}
Experimental inaccessibility of almost all rare decay channels and asymmetry parameters have in the past led to a lot of skepticism about the viability of both theoretical and experimental study of charm in the same glory as granted to beauty and strange physics.   However, that indirect CP violation in charm is ripe for ND intervention has been increasing finding favour amongst flavour physicists. This has been fueled primarily by the conclusive observation of charm oscillations along with a realization that SM in all probabilities cannot account for the same.

Armed with this knowledge, we have proceeded to establish a generalized flavour structure, the LHT-like flavour structure, well tested in $B$, $K$ and $D$ physics,  as a contender for showing effects in two-, three- and four-body decay channels in charm decays. Although, there is little hope that such ND can enhance direct CP violation, indirect CP violation can show up with very unique features, as a very small effect but with sizable CP impurities in the neutral $D$ mass eigenstates. We have also proceeded to comment on a different class of models with non-minimal Higgs sectors which, unlike the LHT-like models, can bring in both direct and indirect CP violation. What effects they can leave on the kind of charm physics discussed in this work is yet to be established concretely. However, it goes without saying that both kinds of ND discussed above can saturate the current experimental bounds set on CP violation in charm.

As we have emphasized in our previous works on rare charm decays within a LHT-like flavour structure, our conclusions are more general than the model that we have tested it in. The same is true for the current work. In addition, decays with higher multiplicity in their final state open up access to more experimental observables although they are theoretically more complicated to deal with than decays with lower multiplicity in the final state. Even with the advent of Super B factories and the LHC where copious numbers of $D$ mesons are produced, tiny branching fractions and large backgrounds, specially for the case of hadron colliders remain obstacles that will have to be tackled in the years to come. 

Despite all the effort that is necessary to glean experimentally observable effects in the $D$ meson system, it is already showing promises of unique effects that have hitherto been unseen in both the $B$ and $K$ systems which have been studied extremely well both theoretically and experimentally. Along with this, the fact that charm lies ripe for the intervention of ND should be sufficient to keep some intrigued souls hard at its heels. 

\section{Acknowledgement}
We would like to thank Lalit Sehgal for comments on his very detailed work that we have cited here and Olivier Pene for some discussions during the early stages of this work. This work was supported by the NSF under Grant No. PHY-0807959.

%

\fancyhf{}
 
\lhead{\uppercase{References}}

\cfoot{\thepage}


\addcontentsline{toc}{section}{References}

\begin{thebibliography}{99}

\bibitem{HYCHENG}
Y. H. Ahn, H-Y Cheng and S. Oh, {\it Wolfenstein Parametrization at Higher Order: Seeming Discrepancies and Their Resolution.} Phys. Lett. {\bf B 703} (2011) 571. 

\bibitem{LHCb1}
LHCb Collaboration: R. Aaij {\em et al.}, {\it Evidence for CP violation in time-integrated $D^0\to h^+h^-$ decay rates}. [arXiv:1112.0938]. 

\bibitem{MIRANDA}
I. Bediaga {\em et al.}, {\it On a CP anisotropy measurement in the Dalitz plot}. Phys.Rev. {\bf D 80} (2009) 096006.

\bibitem{DtoHH}
I. I. Bigi, A. Paul and S. Recksiegel., {\it Conclusions from CDF Results on CP Violation in 
$D^0\to\pi^+\pi^-, K^+K^-$ and Future Task.} JHEP{\bf06} (2011) 089.

\bibitem{CDFDtoHH}CDF Collaboration: T. Aaltonen {\it et al.}, {\it Measurement of CP--violating asymmetries in $D^0\to \pi^+\pi^-$ and $D^0\to K^+K^-$ at CDF.} [arXiv:1111.5023].

\bibitem{HFAGCHARM}HFAG Collaboration, D. Asner et. al., {\it Averages of b-hadron, c-hadron, and $\tau$-lepton Properties}. [arXiv:1010.1589].

 \bibitem{SW}  L. M. Sehgal, M. Wanninger, {\it CP Violation in the Decay $K_L\to\pi^+\pi^-e^+e^-$}. Phys. Rev. {\bf D 46} (1992) 1035; {\it Erratum:} Phys. Rev. {\bf D 46} (1992) 5209.
 
 \bibitem{BK}B. Kayser, {\it Kinematically nontrivial CP violation in beauty decay}. Nucl. Phys. {\bf B (Proc. Suppl.) 13} (1990) 487.
 
 \bibitem{VAR98}I. I. Bigi, in: ``Heavy Flavour Physics: a Probe of Nature's Grand Design", Proc. of the International School of Physics ``Enrico Fermi", Course CXXXVII, IOS Press, (1998) p. 645. [arXiv:hep-ph/9712475], p. 56/57.

\bibitem{DUAL}I. I. Bigi, N. G. Uraltsev, {\it $D^0 - \bar{D}^0$ Oscillations as a Probe of Quark-Hadron Duality}. Nucl. Phys. {\bf B 592} (2001) 92; S. Bianco {et al.}, {\it A Cicerone for the Physics of Charm}. Riv. Nuovo Cim., {\bf 26 N7-8} (2003) p. 151 - 153. 

\bibitem{LENZ}M. Bobrowski et. al., {\it How large can the SM contribution to CP violation in $D^0-\bar{D}^0$ mixing be?}. JHEP {\bf 03} (2010) 009; A. Lenz and M. Bobrowski, {\it Standard Model Predictions for $D^0$-oscillations and CP-violation}. [arXiv:1011.5608].

 \bibitem{FALK}A. Falk et al.,{\it $SU(3)$ Breaking and D0-D0bar Mixing}. Phys. Rev. {\bf D 65} (2002) 054034; {\it The $D^0 - \bar{D}^0$ mass difference from a dispersion relation}. Phys. Rev. {\bf D 69} (2004) 114021. 

 \bibitem{DKdual} I. I. Bigi et al., { \it CP Violation in $D^0-\bar{D}^0$ oscillations: general considerations and applications to the Littlest Higgs model with T parity}. JHEP {\bf 07} (2009) 097.

\bibitem{BLUM} 
K. Blum et al., {\it Combining $K^0-\bar{K}^ 0$ Mixing and $D^0-\bar{D}^ 0$ Mixing to Constrain the Flavor Structure of New Physics}. Phys. Rev. Lett. {\bf 102} (2009) 211802. 

\bibitem{DMB} 
M. Blanke et al. {\it Littlest Higgs model with T-parity confronting the new data on $D^0-\bar{D}^0$ mixing}. Phys. Lett. {\bf B 657} (2007) 81.

\bibitem{SvL} L.M. Sehgal and J. van Leusen, {\it Time evolution of decay spectrum in $K^0, \bar{K}^0 \to\pi^+\pi^-e^+e^-$}. Phys. Lett. {\bf B 489} (2000) 300.

\bibitem{EPR} A. Einstein, B. Podolsky and N. Rosen, {\it Can quantum-mechanical description of physical reality be considered complete?} Phys. Rev {\bf 47} (1935) 777.


\bibitem{HS} P. Heiliger and L. M. Sehgal, {\it Direct and indirect CP violation in the decay $K_L\to \pi^+\pi^-e^+e^-$}. Phys. Rev. {\bf D 48} (1993) 4146; {\it Erratum:} Phys. Rev. {\bf D 60} (1999) 079902.


\bibitem{PBR2} A. Paul, I. I. Bigi, S. Recksiegel, {\it On $D\to X_u l^+ l^-$ within the Standard  Model and Frameworks like the littlest Higgs model with T Parity}. Phys. Rev. {\bf D 83} (2011) 114006.

\bibitem{InamiLim}
T. Inami and C. Lim, {\it Effects of Superheavy Quarks and Leptons in Low-Energy Weak Processes $K_L\to\mu\bar{\mu}$, $K^+\to\pi^+\nu\bar{\nu}$ and $K^0\leftrightarrow\bar{K}^0$}, Prog. Theor. Phys. {\bf 65 (1)} (1981)  297.

\bibitem{Burd}
G. Burdman et al., {\it Rare Charm Decays in the Standard Model and Beyond}, Phys. Rev. {\bf D 66} (2002) 014009.

\bibitem{BuBu}G. Buchalla and A. J. Buras, {\it The rare decays $K^+\to\pi^+\nu\bar{nu}$ and $K_L\to\mu^+\mu^-$  beyond leading logarithms.} Nucl. Phys. {\bf B 412} (1994) 106.

\bibitem{GR} M. Gronau and J. L. Rosner, {\it Triple product asymmetries in $K$, $D_{(s)}$ and $B_{(s)}$ decays.} Phys. Rev. {\bf D 84} (2011) 096013.


\bibitem{LH}N. Arkani-Hamed, A. G. Cohen, and H. Georgi, {\it (De)constructing dimensions}. Phys. Rev. Lett. {\bf 86} (2001) 4757Ð4761; N. Arkani-Hamed, A. G. Cohen, and H. Georgi, {\it Electroweak symmetry breaking from dimensional deconstruction}. Phys. Lett. {\bf B 513} (2001) 232Ð240; N. Arkani-Hamed et al., {\it The minimal moose for a little Higgs}. JHEP {\bf 08} (2002) 21; N. Arkani-Hamed et al., {\it Phenomenology of Electroweak Symmetry Breaking from Theory Space}. JHEP {\bf 08} (2002) 020.

\bibitem{SimHiggs} D. E. Kaplan and M. Schmaltz, {\it The little Higgs from a simple group}. JHEP {\bf 10} (2003) 039; M. Schmaltz, {\it The simplest little Higgs}. JHEP {\bf 08} (2004) 056.

\bibitem{Perel} M. Perelstein, {\it Little Higgs models and their phenomenology}. Prog. in Part. and Nucl. Phys. {\bf 58} (2007) 247.

\bibitem{Han}T. Han, H. E. Logan, B McElrath and L-T Wang, {\it Phenomenology of the little Higgs model}. Phys. Rev. {\bf D 67} (2003) 095004.


\bibitem{CL}H.-C. Cheng and I. Low, {\it TeV symmetry and the little hierarchy problem}. JHEP {\bf 09} (2003) 051; H.-C. Cheng and I. Low, {\it Little hierarchy, little Higgses, and a little symmetry}. JHEP {\bf 08} (2004) 061. 

\bibitem{Low}I. Low,  {\it T parity and the littlest Higgs}. JHEP {\bf 10} (2004) 067.

\bibitem{L2H}N. Arkani-Hamed, A. G. Cohen, E. Katz and A. E. Nelson, {\it The littlest Higgs}. JHEP {\bf 07} (2002) 034.

\bibitem{LHTRev}J. Hubisz and P. Meade. {\it Phenomenology of the littlest Higgs model with T parity}. Phys. Rev. {\bf D 71} (2005) 035016.

\bibitem{VHDVHUCKM}J. Hubisz, S. J. Lee and G. Paz, {\it The flavor of a little Higgs with T parity}. JHEP {\bf 06} (2006) 041.

\bibitem{angph}M. Blanke et al., {\it Another look at the flavour structure of the Littlest Higgs model with T parity}.  Physics Letters {\bf B 646} (2007) 253.

\bibitem{LHTBK}M. Blanke et al., {\it Particle-Antiparticle Mixing\ldots}. JHEP {\bf 12} (2006) 003; M. Blanke et al., {\it Correlations between $\epsilon'/\epsilon$ and rare K decays in the Littlest Higgs model with T parity}. JHEP {\bf 06} (2007) 082;   M. Blanke and A. J. Buras, {\it A Guide to Flavour Changing Neutral Currents in the Littlest Higgs Model with T parity}. Acta Phys. Polon. {\bf B 38} (2007) 2923; A. J. Buras and C. Tarantino, {\it Quark and Lepton Flavour Physics in the Littlest Higgs Model with T parity}. Proceedings of the Workshop {\bf CKM2006}. [arXiv:hep-ph/0702202] (2007); M. Blanke et al., {\it Rare and CP-violating K and B decays in the Littlest Higgs model with T parity}. JHEP {\bf 01} (2007) 066; M. Blanke et al., {\it The Littlest Higgs Model with T parity Facing CP-Violation in $B_s-\bar{B}_s$ Mixing}. [arXiv:0805.4393v2]; M. Blanke et al., {\it FCNC Processes in the Littlest Higgs Model with T parity: an Update}. Acta Phys. Polon. {\bf B 41} (2010) 657.

 \bibitem{PBR1}
A. Paul, I. I. Bigi, S. Recksiegel, {\em $D^0 \to \gamma \gamma$ and $D^0 \to \mu^+ \mu^-$ -- Rates on an Unlikely Impact of the Littlest Higgs Model with T Parity}, Phys. Rev. {\bf D 82} (2010) 094006; {\it Erratum:} Phys. Rev. {\bf D 83} (2011)  019901.

\bibitem{PichA2HDM}A. Pich and P. Tuz\'on, {\it Yukawa alignment in the two-Higgs-doublet model}. Phys. Rev. {\bf D 80} (2009) 091702(R); A. Pich, {\it Flavour constraints on multi-Higgs-doublet models: Yukawa alignment}. Nucl. Phys. {\bf B (Proc. Suppl.) 209} (2010) 182.

\bibitem{PS}S. Pakvasa and H. Sugawara, {\it Discrete Symmetry and Cabibbo Angle}. Phys. Lett. {\bf B 73} (1978) 61.


\end{thebibliography}
\end{document}